\documentclass[english]{article}
\usepackage[T1]{fontenc}
\usepackage[latin9]{inputenc}
\usepackage{geometry}
\usepackage{float}
\usepackage{units}
\usepackage{mathrsfs}
\usepackage{amsmath}
\usepackage{amssymb}
\usepackage{graphicx}

\makeatletter

\floatstyle{ruled}
\newfloat{algorithm}{tbp}{loa}
\providecommand{\algorithmname}{Algorithm}
\floatname{algorithm}{\protect\algorithmname}

\newenvironment{lyxcode}
	{\par\begin{list}{}{
		\setlength{\rightmargin}{\leftmargin}
		\setlength{\listparindent}{0pt}
		\raggedright
		\setlength{\itemsep}{0pt}
		\setlength{\parsep}{0pt}
		\normalfont\ttfamily}%
	 \item[]}
	{\end{list}}

\usepackage[bottom]{footmisc}
\usepackage{natbib}

\title{ \textbf{Optimal quadratic binding for relational reasoning\\ in vector symbolic
neural architectures}}
\author{Naoki Hiratani$^{1}$\thanks{n.hiratani@gmail.com},\, Haim Sompolinsky$^{1,2}$}
\date{}

\usepackage{babel}

\begin{document}
\maketitle
\begin{center}
$^{1}$Center for Brain Science, Harvard University, Cambridge MA 02138, USA 

$^{2}$Edmond and Lily Safra Center for Brain Sciences, Hebrew University,
Jerusalem 91904, Israel
\par\end{center}
\begin{abstract}
Binding operation is fundamental to many cognitive processes, such as cognitive map formation, relational reasoning, and language comprehension.
In these processes, two different modalities, such as location and objects, events and their contextual cues, and words and their roles, need to be bound together, but little is known about the underlying
neural mechanisms. 
Previous works introduced a binding model based on quadratic functions of bound pairs, followed by vector summation of multiple pairs. 
Based on this framework, we address following questions: Which classes of quadratic matrices are optimal for decoding relational structures? And what is the resultant accuracy? 
We introduce a new class of binding matrices based on a matrix representation of octonion algebra, an eight-dimensional extension of complex numbers. We show that these matrices enable a more accurate unbinding than previously known methods when a small number of pairs are present. 
Moreover, numerical optimization of a binding operator converges to this octonion binding. 
We also show that when there are a large number of bound pairs, however, a random quadratic binding performs as well as the octonion and previously-proposed binding methods. 
This study thus provides new insight into potential neural mechanisms of binding operations in the brain.
\end{abstract}
\section{Introduction}

In many cognitive tasks, the brain has to construct a compositional
representation by binding various properties of things like objects,
events, or words. However, little is known about how the brain solves
this binding problem \citep{feldman2013neural}. For example, the scene
depicted in Figure 1A is decomposed into a set of object-location
pairs as
\begin{equation}
[\text{scene}]=\left\{ \left(\text{pink-cube},\text{left}\right),\left(\text{green-pyramid},\text{middle}\right),\left(\text{red-cylinder},\text{\text{right}}\right)\right\} .
\end{equation}
This compositional representation of the object-location pairs is
crucial for scene understanding. For instance, by having this representation
in your working memory, you can answer questions like ``what is the
left-most object?'' (answer: pink-cube), ``what is the position
of the red-cylinder?'' (answer: right) from your memory. However,
it remains elusive how the brain binds neural representations of objects
and locations and creates a compositional representation. Similarly,
in the context of natural language processing, a sentence is interpreted
as a set of word-position pairs:
\begin{equation}
[\text{\textquotedblleft Man bites dog}"]=\left\{ \left(\text{\textquotedblleft man}",1\right),\left(\text{\textquotedblleft bites}",2\right),\left(\text{\textquotedblleft dog}",3\right)\right\} .
\end{equation}
Here, the syntactic position information paired with the words differentiate
the sentence ``man bites dog'' from ``dog bites man'', implying
that the binding of words and their syntactic positions is essential
for language processing. Similar compositional representations are
also suggested to be essential for relational inference, object-based
navigation, and episodic memory formation \citep{eliasmith2012large,whittington2020tolman}.
Moreover, the binding problem is also an important topic in machine
learning literature \citep{greff2020binding}, particularly in knowledge
graph construction \citep{socher2013reasoning,nickel2016holographic},
and relational reasoning \citep{johnson2017clevr,santoro2017simple}. 

\begin{figure}
\begin{centering}
\includegraphics[scale=1.1]{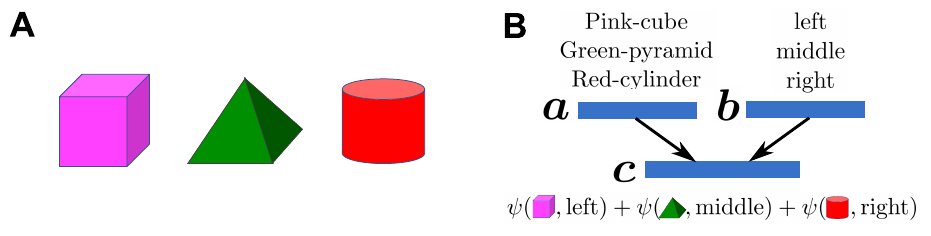}\caption{Schematics of the binding problem. \textbf{A)} A scene with three objects:
pink-cube (left), green-pyramid (middle), and red-cylinder (right).
\textbf{B)} Representation of the scene in VSA (vector symbolic architecture).
Vector representation of objects ($\boldsymbol{a}$) and their positions
($\boldsymbol{b})$ are combined into a compositional representation
of the entire scene ($\boldsymbol{c}$). }
\rule{\textwidth}{0.25pt}
\par\end{centering}
\end{figure}

Mathematically speaking, this is a problem of vector representation
construction. Let us consider a vector representation of a set of
pairs $S=\left\{ \left(\boldsymbol{a}_{\mu},\boldsymbol{b}_{\mu}\right)\right\} _{\mu=1}^{L}$,
where $L$ is the number of the pairs, and $\boldsymbol{a}$ and $\boldsymbol{b}$
are $N$-dimensional vectors. For instance, in the case of the scene
recognition depicted in Figure 1A, $\boldsymbol{a}_{1}$ is a vector
representation of ``pink-cube'', while $\boldsymbol{b}_{1}$ is
a representation of the position ``left'', and so on (Fig. 1B).
In the brain, the number of neurons recruited for a representation
of an object or its position is expectedly large, whereas previous
human studies indicate that the number of pairs, $L$, the brain can
hold in the short-term memory is less than ten \citep{miller1956magical,cowan2001magical}.
Therefore, we will mainly focus on the parameter regime where $1\lesssim L\ll N$
is satisfied. 

Previous works proposed the vector symbolic architecture (VSA) as
a biological-plausible solution for the binding problem \citep{smolensky1990tensor,plate1995holographic,gayler2004vector,kanerva2009hyperdimensional}.
In particular, VSA is capable of instantaneous construction of compositional
structures essential for linguistic processing \citep{gayler2004vector}.
In the VSA framework, a vector representation of a set $S=\left\{ \left(\boldsymbol{a}_{\mu},\boldsymbol{b}_{\mu}\right)\right\} _{\mu=1}^{L}$
is constructed by:
\begin{equation}
\begin{array}{c}
\boldsymbol{c}_{\mu}=\psi\left(\boldsymbol{a}_{\mu},\boldsymbol{b}_{\mu}\right)\quad:\text{binding}\\
\boldsymbol{c}=\sum_{i=1}^{L}\boldsymbol{c}_{\mu}\quad:\text{bundling}
\end{array}
\end{equation}
where $\boldsymbol{c}$ is a $N_{c}$-dimensional vector, and $\psi$
is a non-linear mapping $\psi:\mathbb{R}^{N}\times\mathbb{R}^{N}\to\mathbb{R}^{N_{c}}$.
This means that we first create a representation of a pair $\left(\boldsymbol{a}_{\mu},\boldsymbol{b}_{\mu}\right)$
by a non-linear mapping $\psi\left(\boldsymbol{a}_{\mu},\boldsymbol{b}_{\mu}\right)$,
then generate a representation of the set $S$ by summing up the representation
of the pairs (Fig. 1B). By constructing a representation of a set
by taking the sum over pairs, the length of vector $\boldsymbol{c}$
stays constant regardless of the cardinality of the set $L$. This
is a desirable property when we consider population coding by a fixed
number of neurons. However, it also causes interference between different
pairs, as we will see. 

In this paper, we study how we should choose the binding operator
$\psi$. The answer depends on the objective, but it is often desirable
for $\boldsymbol{c}$ to maximize the unbinding performance. In other
words, we should be able to retrieve $\boldsymbol{a}_{\mu}$ from
$\boldsymbol{c}$ using $\boldsymbol{b}_{\mu}$ as a query, and vice
versa. In our example (Fig. 1A), representation of the scene, $\boldsymbol{c}$,
should enable us to answer a question like ``what is the left-most
object?'' (answer: pink-cube) or ``what is the position of the red-cylinder?''
(answer: right). 

Previous works introduced two binding mechanisms for VSA architecture,
the holographic reduced representation (HRR) \citep{plate1995holographic}
and the tensor product representation \citep{smolensky1990tensor},
among others \citep{kanerva1997fully,gallant2013representing,gosmann2019vector,frady2020variable}.
HRR is noisy, but the size of composition $\boldsymbol{c}$ is the
same with its elements $\boldsymbol{a},\boldsymbol{b}$ (i.e., $N_{c}=N$).
On the other hand, the tensor product representation is more accurate,
but it requires $N_{c}=N^{2}$ neurons for representing a composition
(see Appendix \ref{subsec:HRR} and \ref{subsec:TPR} for the details
of the two binding methods). Though their properties have been studied
previously \citep{plate1997common,schlegel2020comparison,steinberg2022associative},
it remains elusive if HRR and the tensor product representation are
the optimal binding under $N_{c}=N$ and $N_{c}=N^{2}$ respectively.
Moreover, little is known on how we should construct a binding operator
under various composition sizes $N_{c}$ and how the minimum achievable
error scales with the number of bound pairs $L$. Below, we address
these questions under a quadratic parameterization of the binding
operators. We found that at $L\sim O(1)$, there is a novel binding
algorithm based on a matrix representation of the octonion algebra
that significantly outperforms HRR and its extension. We also show
that when $L\gg1$ and $N_{c}\ll N^{2}$, there is no quadratic binding
method that significantly outperforms a random binding method. 

\section{Quadratic binding}

Below, we introduce a specific class of binding operators that has
a quadratic form. More specifically, using an $N\times N\times N_{c}$
tensor $P$, we define the $k$-th element of a representation of
set $S=\left\{ \left(\boldsymbol{a}_{\mu},\boldsymbol{b}_{\mu}\right)\right\} _{\mu=1}^{L}$
as 
\begin{equation}
c_{k}=\sum_{\mu=1}^{L}\sum_{i=1}^{N}\sum_{j=1}^{N}P_{ijk}a_{i}^{\mu}b_{j}^{\mu},\label{eq_def_qbind}
\end{equation}
for $k=1,...,N_{c}$, where $a_{i}^{\mu}$ is the $i$-th element
of vector $\boldsymbol{a}_{\mu}$. There are several motivations for
why we consider this quadratic parameterization. First, assuming that
the norm of the vectors is constant ($\left\Vert \boldsymbol{a}\right\Vert ^{2}=\left\Vert \boldsymbol{b}\right\Vert ^{2}=N$),
many previously proposed binding operators, such as HRR and the tensor
product representation are written as examples of quadratic binding.
For instance, if we set $P_{ijk}=\delta_{[i+j]_{N},k}$ with $[i+j]_{N}\equiv i+j\;(\textrm{mod. }N)$,
we recover HRR, $c_{k}=\sum_{\mu}\sum_{i}a_{i}^{\mu}b_{[k-i]_{N}}^{\mu}$
(see Appendix \ref{subsec:HRR}). Moreover, when $\boldsymbol{a}$
and $\boldsymbol{b}$ are random Gaussian variables, the quadratic
parameterization should be enough to capture the statistical relationship
between $\boldsymbol{a}$ and $\boldsymbol{b}$. Thirdly, this formulation
is simple enough to be biologically plausible, though the biological
substrates for the multiplication are not yet fully understood.

From this vector representation $\boldsymbol{c}$, we consider unbinding
of a vector using its bound pair as a query. For example, to answer
the question ``what is the left-most object?'' from a vector representation
of the scene depicted in Fig. 1A, we need to unbind ``pink-cube''
from representation $\boldsymbol{c}$ by using the position ``left''
as a query. Here, we also restrict this unbinding operation onto a
quadratic form. Unbinding of $\boldsymbol{a}_{1}$ with a query $\boldsymbol{b}_{1}$
is defined by
\begin{equation}
\widehat{a}_{i}^{1}=\sum_{j=1}^{N}\sum_{k=1}^{N_{c}}Q_{ijk}b_{j}^{1}c_{k},\label{eq_def_qunbind_a}
\end{equation}
where $Q$ is an $N\times N\times N_{c}$ tensor. Similarly, using
an $N\times N\times N_{c}$ tensor $R$, unbinding of $\boldsymbol{b}_{1}$
with $\boldsymbol{a}_{1}$ is defined by
\begin{equation}
\widehat{b}_{j}^{1}=\sum_{i=1}^{N}\sum_{k=1}^{N_{c}}R_{ijk}a_{i}^{1}c_{k}.\label{eq_def_qunbind_b}
\end{equation}

Our objective is to find a set of tensors $P,Q,R$ that achieves the
best unbinding performance. Using the mean-squared error as the loss,
we define the unbinding error of $\boldsymbol{a}_{1}$ and $\boldsymbol{b}_{1}$
as 
\begin{equation}
\ell_{a}\left(P,Q\right)\equiv\frac{1}{N}\left\langle \left\Vert \boldsymbol{a}_{1}-\widehat{\boldsymbol{a}}_{1}\right\Vert ^{2}\right\rangle _{p(S)},\label{eq_def_la}
\end{equation}
and
\begin{equation}
\ell_{b}\left(P,R\right)\equiv\frac{1}{N}\left\langle \left\Vert \boldsymbol{b}_{1}-\widehat{\boldsymbol{b}}_{1}\right\Vert ^{2}\right\rangle _{p(S)}.\label{eq_def_lb}
\end{equation}
Below, we consider the case when $\boldsymbol{a}$ and $\boldsymbol{b}$
are sampled from an i.i.d Gaussian distribution $N(0,I_{N})$. This
assumption is introduced partially for analytical tractability, but
we would also expect the input vectors $\boldsymbol{a}$ and $\boldsymbol{b}$
to be whitened in the preprocessing. 

Inserting Eq. \ref{eq_def_qbind} and Eq. \ref{eq_def_qunbind_a}
into the loss $\ell_{a}$ (Eq. \ref{eq_def_la}), we get 
\begin{equation}
\ell_{a}=\frac{1}{N}\left\langle \sum_{i=1}^{N}\left(a_{i}^{1}-\sum_{\mu=1}^{L}\sum_{j=1}^{N}\sum_{l=1}^{N}\sum_{m=1}^{N}\left[\sum_{k=1}^{N_{c}}Q_{ijk}P_{lmk}\right]b_{j}^{1}b_{m}^{\mu}a_{l}^{\mu}\right)^{2}\right\rangle _{p(\boldsymbol{a},\boldsymbol{b})}.
\end{equation}
Because the error depends only on the tensor product of $P$ and $Q$
over index $k$, there is an invariance in the choice of $P$ and
$Q$. If we define $\widetilde{Q}_{ijk}=\sum_{n}Q_{ijn}A_{nk}$ and
$\widetilde{P}_{lmk}=\sum_{n}P_{lmn}\left[A^{-1}\right]_{kn}$ with
an $N_{c}\times N_{c}$ invertible matrix $A$, we get $\sum_{k}Q_{ijk}P_{lmk}=\sum_{k}\widetilde{Q}_{ijk}\widetilde{P}_{lmk}$,
indicating that the choice of the optimal $P$ and $Q$ is not unique.
Taking the expectation over $\boldsymbol{a}$ and $\boldsymbol{b}$,
the equation above is rewritten as (see Appendix \ref{subsec:Fixed-point-condition})
\begin{equation}
\ell_{a}=1-\frac{2}{N}\sum_{i}tr\left[P_{i}Q_{i}^{T}\right]+\frac{1}{N}\sum_{i}\sum_{l}\left(\left(tr\left[P_{l}Q_{i}^{T}\right]\right)^{2}+tr\left[P_{l}Q_{i}^{T}\left(L\cdot Q_{i}P_{l}^{T}+P_{l}Q_{i}^{T}\right)\right]\right),\label{eq_la_L1_2}
\end{equation}
where all summations run from $1$ to $N$, and $P_{i}$ and $Q_{i}$
are $N\times N$ matrices corresponding to the $i$-th component of
tensors $P$ and $Q$, respectively:
\begin{equation}
\left[P_{i}\right]_{jk}=P_{ijk},\quad\left[Q_{i}\right]_{jk}=Q_{ijk}.
\end{equation}
Similarly, by taking the expectation over $\boldsymbol{a}$ and $\boldsymbol{b}$,
the decoding loss of $\boldsymbol{b}$ is given as 
\begin{equation}
\ell_{b}=1-\frac{2}{N}\sum_{i}tr\left[P_{i}R_{i}^{T}\right]+\frac{1}{N}\sum_{i}\sum_{l}\left(tr\left[P_{i}R_{i}^{T}R_{l}P_{l}^{T}\right]+tr\left[P_{l}R_{i}^{T}\left(L\cdot R_{i}P_{l}^{T}+R_{l}P_{i}^{T}\right)\right]\right),\label{eq_lb_L1}
\end{equation}
where $\left[R_{i}\right]_{jk}=R_{ijk}$. 

\section{The binding solutions under $L=1$ }

How should we choose binding operator $P$ and unbinding operators
$Q,R$ to minimize the loss $\ell_{a}$ and $\ell_{b}$? Let us start
from a simple scenario where only one pair is bound (i.e., $L=1$),
and the size of the composition $\boldsymbol{c}$ is the same with
its elements $\boldsymbol{a}$ and $\boldsymbol{b}$ (i.e., $N_{c}=N$).
In this scenario, there is actually a trivial non-quadratic lossless
algorithm in which binding and unbinding are performed by $\boldsymbol{c}=\boldsymbol{a}+\boldsymbol{b}$
and $\hat{\boldsymbol{a}}=\boldsymbol{c}-\boldsymbol{b}$, which we
call the sum binding. However, this strategy scales badly to $L>1$
as we will see later. Below, we first investigate the solution numerically
using a fixed-point algorithm, then subsequently, study a sufficient
condition for a local minimum of $\ell_{a}$ and $\ell_{b}$ analytically. 

\subsection{Numerical optimization of the binding tensor}

To investigate the solution space of the quadratic binding operators,
we first optimize the binding operators $P,Q,R$ numerically for both
$\ell_{a}$ and $\ell_{b}$ using a fixed-point algorithm. Taking
the gradient of $\ell_{a}$ with respect to $P_{l}$ and rewriting
this equation in a tensor form, the fixed-point condition is given
as (see Appendix \ref{subsec:numerical_opt} for the details)
\begin{equation}
\frac{\partial\ell_{a}}{\partial P_{l}}=0\Leftrightarrow Q_{ljk}=\sum_{m,n}\Gamma_{[jN+k],[mN+n]}^{q}P_{lmn},\label{eq_dla_dp}
\end{equation}
for $j,k=1,...,N$, where $\Gamma^{q}$ is a $N^{2}\times N^{2}$
matrix defined as 
\begin{equation}
\Gamma_{[jN+k],[mN+n]}^{q}\equiv\sum_{i}\left(\delta_{jm}\sum_{\beta}Q_{i\beta k}Q_{i\beta n}+Q_{ijk}Q_{imn}+Q_{ijn}Q_{imk}\right).
\end{equation}
Therefore, for a given unbinding tensor $Q$, the binding tensor $P$
satisfying $\frac{\partial\ell_{a}}{\partial P_{l}}=0$ is given as
\begin{equation}
\text{Vec}\left[P_{l}\right]=\left(\Gamma^{q}\right)^{-1}\text{Vec}\left[Q_{l}\right],
\end{equation}
where $\text{Vec}\left[P_{l}\right]$ and $\text{Vec}\left[Q_{l}\right]$
are the vector representation of $N\times N$ matrices $P_{l}$ and
$Q_{l}$, respectively. From a similar calculation, for a given binding
tensor $P$, the fixed point of $\ell_{a}$ with respect to $Q$ is
given as
\begin{equation}
\text{Vec}\left[Q_{l}\right]=\left(\Gamma^{pa}\right)^{-1}\text{Vec}\left[P_{l}\right],
\end{equation}
where $\Gamma^{pa}$ is an $N^{2}\times N^{2}$ matrix that only depends
on $P$. Moreover, we can update $P$ and $R$ with respect to $\ell_{b}$
by 
\begin{equation}
\text{Vec}\left[\check{P}_{m}\right]=\left(\Gamma^{r}\right)^{-1}\text{Vec}\left[\check{R}_{m}\right],\quad\text{Vec}\left[\check{R}_{j}\right]=\left(\Gamma^{pb}\right)^{-1}\text{Vec}\left[\check{P}_{j}\right]
\end{equation}
respectively, where $\check{R}_{m}$ and $\check{P}_{m}$ are defined
as $\left[\check{R}_{m}\right]_{l,k}=R_{lmk}$ and $\left[\check{P}_{m}\right]_{l,k}=P_{lmk}$,
and $\Gamma^{r}$ and $\Gamma^{pb}$ are $N^{2}\times N^{2}$ matrices
(defined at Eqs. \ref{eq_def_Gamma_r} and \ref{eq_def_Gamma_pb}
in Appendix \ref{subsec:numerical_opt}). Therefore, we can perform
an iterative optimization of tensors $P,Q,R$ with respect to both
$\ell_{a}$ and $\ell_{b}$ by the fixed-point algorithm described
in Algorithm 1.
\begin{algorithm}
\caption{A fixed-point algorithm of $P,Q,R$ optimization}

\begin{lyxcode}
randomly~initialize~$P,Q,R$

for~t~in~1...T:
\begin{lyxcode}
$\text{Vec}[Q_{i}]=\left(\Gamma^{pa}\right)^{-1}\text{Vec}\left[P_{i}\right]$~for~$i=1,...,N$

$\text{Vec}[P_{i}]=\left(\Gamma^{q}\right)^{-1}\text{Vec}\left[Q_{i}\right]$~~for~$i=1,...,N$

$\text{Vec}[\check{R}_{j}]=\left(\Gamma^{pb}\right)^{-1}\text{Vec}\left[\check{P}_{j}\right]$~for~$j=1,...,N$

$\text{Vec}[\check{P}_{j}]=\left(\Gamma^{r}\right)^{-1}\text{Vec}\left[\check{R}_{j}\right]$~~for~$j=1,...,N$
\end{lyxcode}
\end{lyxcode}
\end{algorithm}

Figures 2A and B describe the learning curves of this algorithm from
ten random initializations. Both decoding errors $\ell_{a}$ and $\ell_{b}$
first plateau around $\ell_{a}\approx0.35$, then converge to $1/5$,
robustly. To evaluate the performance of this optimized binding operator,
we compare it with HRR \citep{plate1995holographic}, a binding method
using the circular convolution (see Appendix \ref{subsec:HRR}). Under
HRR, the binding and the unbinding operators are given as, 
\begin{equation}
P_{ijk}=Q_{ijk}=R_{ijk}=\frac{1}{\sqrt{2(N+1)}}\delta_{[i+j]_{N},k},
\end{equation}
where $[x]_{N}\equiv x\;(\textrm{mod. }N)$, and the decoding error
is $\ell_{a}=\ell_{b}=1/2$ under a large $N$ (gray dashed lines
in Figs. 2A and B). This means that the numerically optimized binding
operator achieves a better decoding performance than HRR under $L=1$. 

\begin{figure}
\centering{}\includegraphics[scale=0.8]{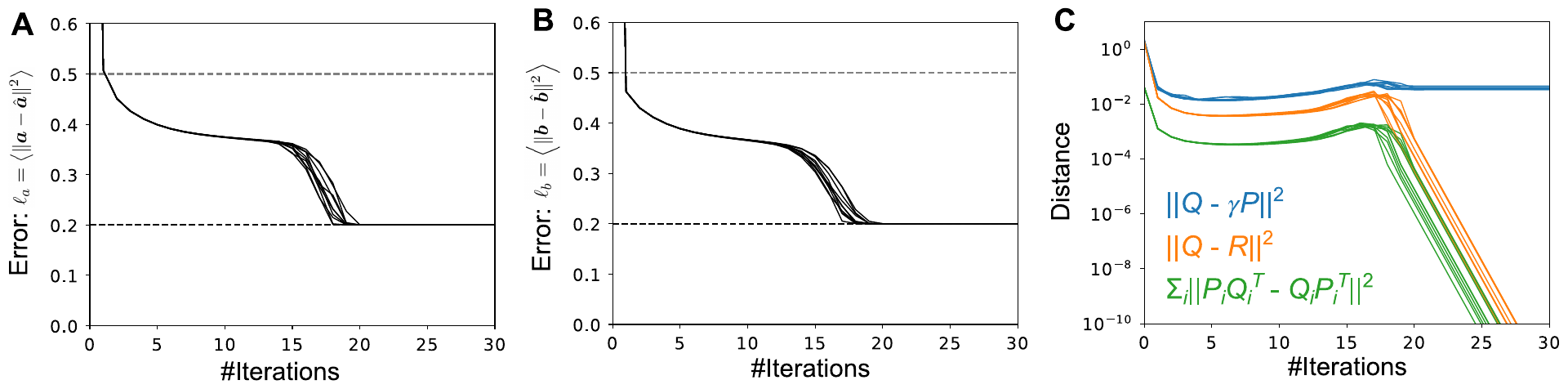}\caption{Iterative optimization of binding operators $P,Q,R$ under $N=48$ and $L=1$. 
\textbf{A)} Learning dynamics of the fixed-point algorithm from random initializations. Here, we evaluated the loss $\ell_{a}$ directly from Eq. \ref{eq_la_L1_2}. Each line represents a learning curve of the decoding error $\ell_{a}$ from a random initialization. The black dashed line represents the error after convergence, and gray dashed line is the error under HRR. 
\textbf{B)} The same as \textbf{A}, but the y-axis
is $\ell_{b}$, not $\ell_{a}$.
\textbf{C)} Change in distance measures during
learning. We defined the distances using the Frobenius norm as $\left\Vert Q-\gamma P\right\Vert ^{2}\equiv\sum_{i,j,k}\left(Q_{ijk}-\gamma P_{ijk}\right)^{2}$
where $\gamma=\tfrac{\max Q}{\max P}$, $\left\Vert Q-R\right\Vert ^{2}\equiv\sum_{i,j,k}\left(Q_{ijk}-R_{ijk}\right)^{2}$
, and $\sum_{i}\left\Vert P_{i}Q_{i}^{T}-Q_{i}P_{i}^{T}\right\Vert ^{2}\equiv\sum_{ijk}\left(\left[P_{i}Q_{i}^{T}\right]_{jk}-\left[P_{i}Q_{i}^{T}\right]_{kj}\right)^{2}$.}
\rule{\textwidth}{0.25pt}
\end{figure}

To see if the numerically optimized binding operators $P,Q,R$ have
some specific structures, we next investigate the values of $P,Q,R$
after learning. Firstly, upon optimization, $Q$ and $R$ converge
to the same values (i.e., $Q=R$; orange lines in Fig. 2C), but $P\neq Q,R$
even under a rescaling (blue lines; here we plotted a normalized distance
$\sum_{ijk}\left(Q_{ijk}-\gamma P_{ijk}\right)^{2}$ with $\gamma=\tfrac{\max Q}{\max P}$).
Moreover, the matrix products $P_{1}Q_{1}^{T},...,P_{N}Q_{N}^{T}$
converge to symmetry matrices after the optimization (green lines). 

Elements of matrix $P_{i}$ look random even after an optimization
(Fig. 3A; here we plotted $P_{1}$,...,$P_{5}$ out of $N=48$ matrices
$P_{1}$,...,$P_{48}$) potentially due to an invariance in the solution
space, and the same is true for the elements of $Q_{i}$ (Fig. 3B).
To untangle the invariance and extract the hidden structure in $P,Q,R$,
we process the tensors in the following way: 
\begin{enumerate}
\item Because $P_{1}Q_{1}^{T}$ converges to a symmetric matrix, we can decompose it as $P_{1}Q_{1}^{T}=U_{1}\Sigma_{1}U_{1}^{T}$ where $U_{1}$ is an $N\times N$ orthogonal matrix, and $\Sigma_{1}$ is an $N\times N$
diagonal matrix. 
\item We introduce an $N\times N$ matrix $A_{1}$ by $A_{1}\equiv P_{1}^{-1}U_{1}\Sigma_{1}^{1/2}$. 
\item We transform the binding matrices $P_{i}$ by $\bar{P}_{i}=U_{1}^{T}P_{i}A_{1}$
for $i=1,...,N$.
\end{enumerate}
This transformation cancels out the invariance in the choice of $P$
and $Q$, and maps $P_{i}$ onto the space where $\bar{P}_{1}$ is
a diagonal matrix ($\bar{P}_{1}=\Sigma_{1}^{1/2}$). After this preprocessing,
we found an $8\times8$ block structure in all $\bar{P}_{i}$ (Fig.
3C; we plotted $\bar{P}_{1}$,...,$\bar{P}_{5}$ out of $\bar{P}_{1}$,...,$\bar{P}_{48}$
as before). Note that, there is no constraint that enforces the $8\times8$
structure in the learning algorithm nor the data processing, except
that $N=48$ is a multiple of eight. Similarly, by preprocessing $Q_{i}$
by $\bar{Q}_{i}=U_{1}^{T}Q_{i}B_{1}$ with $B_{1}=Q_{1}^{-1}U_{1}\Sigma^{1/2}$,
we recover the $8\times8$ block structure (Fig. 3D). Moreover, $\bar{P}_{i}=\bar{Q_{i}}$
is satisfied for all $i=1,...,N$ (compare Figs. 3C and 3D). Notably,
$\bar{P}_{i}\bar{Q}_{i}^{T}$ is rewritten as 
\begin{equation}
\bar{P}_{i}\bar{Q}_{i}^{T}=U_{1}^{T}P_{i}P_{1}^{-1}U_{1}\Sigma_{1}U_{1}^{T}\left(Q_{1}^{T}\right)^{-1}Q_{i}U_{1}^{T}=U_{1}^{T}P_{i}Q_{i}^{T}U_{1}.
\end{equation}
Therefore, under the transformation $\left\{ P_{i},Q_{i}\right\} \to\left\{ \bar{P_{i}},\bar{Q}_{i}\right\} $,
the loss $\ell_{a}$ (Eq. \ref{eq_la_L1_2}) is preserved. This means
that, up to a linear transformation with an orthogonal matrix, these
numerically optimized operators are symmetric ($Q=R$ and $\bar{P}=\bar{Q}$)
and have a hidden $8\times8$ block structure. In the rest of the
section, we discuss why we see the $8\times8$ structure in the optimized
binding operators from an algebraic perspective.

\begin{figure}
\begin{centering}
\includegraphics{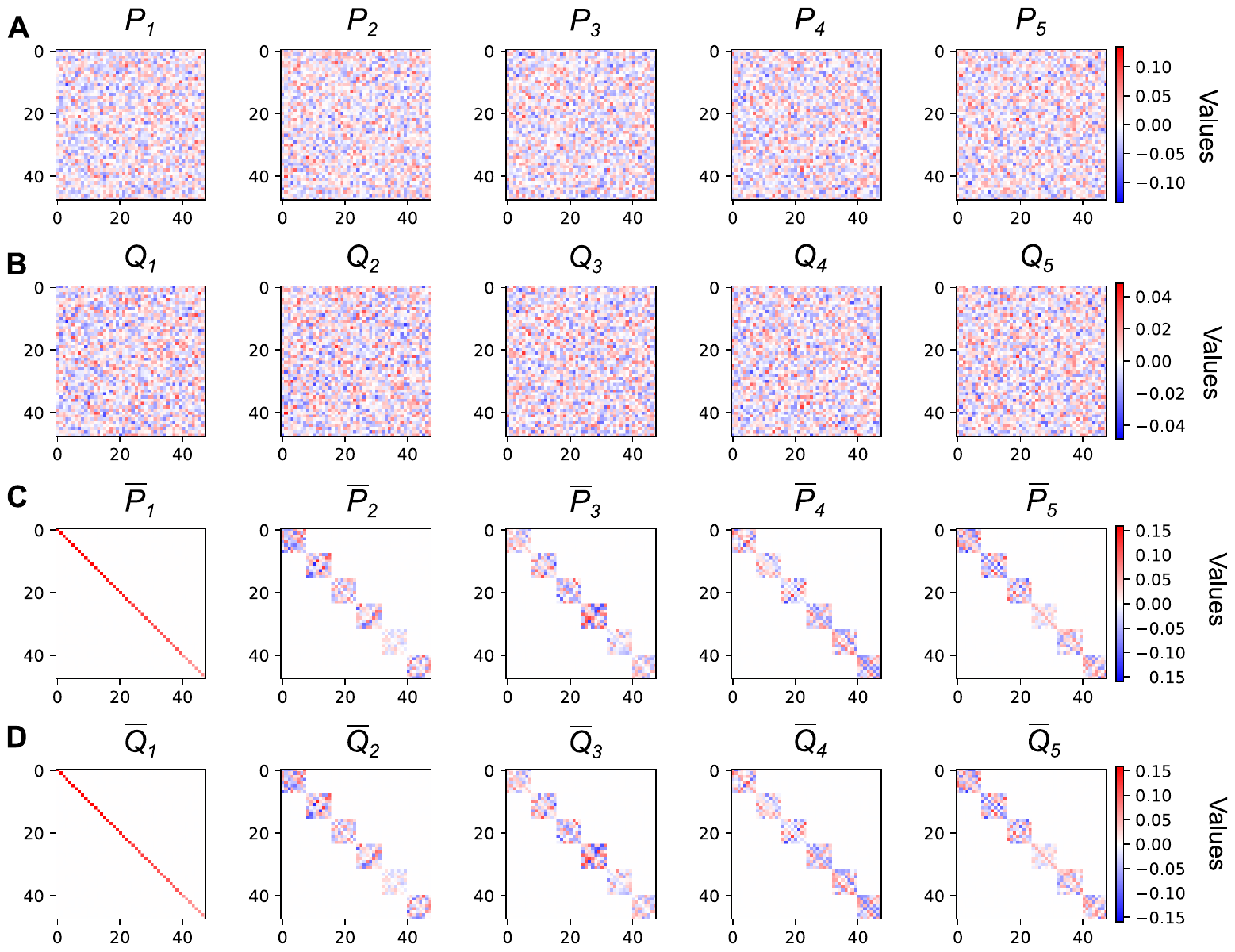}\caption{Binding matrices obtained after the convergence from one random initialization ($N=48$, $L=1$).
\textbf{A)} $P_{1},...,P_{5}$ (of $P_{1},...,P_{48}$) after 100 iterations of the numerical optimization. 
\textbf{B)} $Q_{1},...,Q_{5}$ after the numerical optimization. 
\textbf{C, D)} The same as the binding matrices depicted in panels \textbf{A} and \textbf{B} respectively, but transformed into a space
where $\bar{P}_{1}$ is diagonal. }
\rule{\textwidth}{0.25pt}
\par\end{centering}
\end{figure}

\subsection{Composition algebra-based solution for the quadratic binding problem}

Our numerical optimization indicates that there is a non-trivial binding
method with a $8\times8$ block diagonal structure that outperforms
a previously proposed method. To understand the origin of the $8\times8$
block structure, we next analytically study a sufficient condition
for a local minimum of both $\ell_{a}$ and $\ell_{b}$. Below, we
introduce $P=Q=R$ constraint for the binding and unbinding tensors.
This is motivated by the symmetric structure we saw in the numerical
optimization (Figs. 2C and 3CD). Note that, two popular binding methods,
HRR and the tensor product representation, also satisfy this $P=Q=R$
constraint (see Appendix \ref{sec:Tensor-Holographic-bindings}).
Taking the gradient of $\ell_{a}$ under the symmetric constraint
$P=Q$, the fixed-point condition is given as
\begin{equation}
P_{i}=\sum_{l=1}^{N}\left(tr\left[P_{l}P_{i}^{T}\right]I_{N}+P_{l}P_{i}^{T}+P_{i}P_{l}^{T}\right)P_{l}\label{eq_dla_dp_sym}
\end{equation}
for $i=1,...,N$, where $I_{N}$ is the size $N$ identity matrix.
Similarly, introducing $P=R$ constraint, the fixed-point condition
for $\ell_{b}$ is written as 
\begin{equation}
P_{i}=\sum_{l=1}^{N}\left(P_{i}P_{l}^{T}P_{l}+P_{l}P_{i}^{T}P_{l}+P_{l}P_{l}^{T}P_{i}\right).\label{eq_dlb_dp_sym}
\end{equation}

What kind of $P=\left\{ P_{1},P_{2},...,P_{N}\right\} $ satisfies
Eqs. \ref{eq_dla_dp_sym} and \ref{eq_dlb_dp_sym}? A sufficient condition
for both Eqs. \ref{eq_dla_dp_sym} and \ref{eq_dlb_dp_sym} is
\begin{equation}
P_{l}P_{i}^{T}+P_{i}P_{l}^{T}=2\lambda\delta_{il}I_{N},\label{eq_hurwitz_mat_eq}
\end{equation}
for all $i,l=1,...,N$, with the scaling factor $\lambda=\frac{1}{N+2}$
(see Appendix \ref{subsec:sufficiency}). This set of equations is
known as the Hurwitz matrix equations. It has been proved that, there
exists a family of $N$ matrices of size $N\times N$ that satisfies
Eq. \ref{eq_hurwitz_mat_eq} only if $N=1,2,4,8$, and a solution
is given by a real matrix representation of the composition algebra
of dimension $N$ \citep{shapiro2011compositions}. This means that,
when binding two vectors $\boldsymbol{a},\boldsymbol{b}$ having the
length $N=1,2,4,8$, you can locally minimize the decoding error by
using a solution of the Hurwitz matrix equations as a binding operator
$P$. 

For instance, when $N=2$, by setting 
\begin{equation}
P_{1}=\frac{1}{2}\left(\begin{array}{cc}
1 & 0\\
0 & 1
\end{array}\right),\;P_{2}=\frac{1}{2}\left(\begin{array}{cc}
0 & 1\\
-1 & 0
\end{array}\right),
\end{equation}
Eq. \ref{eq_hurwitz_mat_eq} is satisfied. Then, from Eq. \ref{eq_def_qbind},
binding of two (real) vectors $\boldsymbol{a}=\left(a_{1},a_{2}\right)$
and $\boldsymbol{b}=\left(b_{1},b_{2}\right)$ becomes 
\begin{equation}
\boldsymbol{c}=\frac{1}{2}\left(\begin{array}{c}
a_{1}b_{1}-a_{2}b_{2}\\
a_{1}b_{2}+a_{2}b_{1}
\end{array}\right),
\end{equation}
and unbinding of $\boldsymbol{a}$ is given as $\hat{\boldsymbol{a}}=\frac{b_{1}^{2}+b_{2}^{2}}{4}\left(a_{1},a_{2}\right)$.
Notably, $P_{1}$ and $P_{2}$ consist of a basis of a matrix representation
of the complex numbers up to a scaling factor. Let us define a projection
$\phi:\mathbb{C}\to\mathtt{\mathbb{R}}^{2}\times\mathbb{R}^{2}$ by
\begin{equation}
\phi\left(x+iy\right)=2\left(xP_{1}+yP_{2}\right)=\left(\begin{array}{cc}
x & y\\
-y & x
\end{array}\right).
\end{equation}
Then, for two complex numbers $a=a_{1}+ia_{2}$ and $b=b_{1}+ib_{2}$,
\begin{equation}
\phi\left(a\right)\phi\left(b\right)=\left(\begin{array}{cc}
a_{1} & a_{2}\\
-a_{2} & a_{1}
\end{array}\right)\left(\begin{array}{cc}
b_{1} & b_{2}\\
-b_{2} & b_{1}
\end{array}\right)=\left(\begin{array}{cc}
a_{1}b_{1}-a_{2}b_{2} & a_{1}b_{2}+a_{2}b_{1}\\
-(a_{1}b_{2}+a_{2}b_{1}) & a_{1}b_{1}-a_{2}b_{2}
\end{array}\right)=\phi\left(ab\right).
\end{equation}
Thus, at $N=2$, a matrix representation of the complex numbers provides
a binding operator $P=\left[P_{1},P_{2}\right]$ that satisfies the
fixed-point conditions (Eqs. \ref{eq_dla_dp_sym} and \ref{eq_dlb_dp_sym}).
Similarly, at $N=4$, we can construct a binding operator $P=\left[P_{1},P_{2},P_{3},P_{4}\right]$
using a matrix representation of the quaternions:
{\footnotesize
\begin{equation}
P_{1}=\frac{1}{\sqrt{6}}\left(\begin{array}{cccc}
1 & 0 & 0 & 0\\
0 & 1 & 0 & 0\\
0 & 0 & 1 & 0\\
0 & 0 & 0 & 1
\end{array}\right),\;P_{2}=\frac{1}{\sqrt{6}}\left(\begin{array}{cccc}
0 & 1 & 0 & 0\\
-1 & 0 & 0 & 0\\
0 & 0 & 0 & 1\\
0 & 0 & -1 & 0
\end{array}\right),\;P_{3}=\frac{1}{\sqrt{6}}\left(\begin{array}{cccc}
0 & 0 & 1 & 0\\
0 & 0 & 0 & -1\\
-1 & 0 & 0 & 0\\
0 & 1 & 0 & 0
\end{array}\right),\;P_{4}=\frac{1}{\sqrt{6}}\left(\begin{array}{cccc}
0 & 0 & 0 & 1\\
0 & 0 & 1 & 0\\
0 & -1 & 0 & 0\\
-1 & 0 & 0 & 0
\end{array}\right).
\end{equation}
}
Under this binding operator, the composition $\boldsymbol{c}$ of
the two vectors $\boldsymbol{a}$ and $\boldsymbol{b}$ is given as
\begin{equation}
\begin{array}{c}
c_{1}=\frac{1}{\sqrt{6}}\left(a_{1}b_{1}+a_{2}b_{2}+a_{3}b_{3}+a_{4}b_{4}\right)\\
c_{2}=\frac{1}{\sqrt{6}}\left(a_{1}b_{2}-a_{2}b_{1}-a_{3}b_{4}+a_{4}b_{3}\right)\\
c_{3}=\frac{1}{\sqrt{6}}\left(a_{1}b_{3}+a_{2}b_{4}-a_{3}b_{1}-a_{4}b_{2}\right)\\
c_{4}=\frac{1}{\sqrt{6}}\left(a_{1}b_{4}-a_{2}b_{3}+a_{3}b_{2}-a_{4}b_{1}\right)
\end{array}
\end{equation}
and using $P$ as the unbinding tensor, from Eq. \ref{eq_def_qunbind_a},
we get 
\begin{equation}
\widehat{\boldsymbol{a}}=\frac{1}{6}\left(b_{1}^{2}+b_{2}^{2}+b_{3}^{2}+b_{4}^{2}\right)\boldsymbol{a}.\label{eq_quarternion_unbinding}
\end{equation}
Therefore, decoding of $\boldsymbol{a}$ is faithful up to a constant
scaling factor under this quaternion-based binding. Similarly, we
can construct a binding operator based on a matrix representation
of the octonions, an extension of the quaternions to $N=8$ dimensional
space (see Appendix \ref{subsec:Octonion-binding}). However, this
is not true for $N>8$, because Eq. \ref{eq_hurwitz_mat_eq} does
not admit a solution. 

\subsection{Sparse $K$-compositional bindings}

Although solutions based on the composition algebra we discussed above
are lossless, they cannot be directly extended to the case when $N>8$
because the Hurwitz matrix equations do not have a solution. Nevertheless,
we can apply this binding in a block-wise manner. Let us define a
family of matrices that satisfies the Hurwitz matrix equations (Eq.
\ref{eq_hurwitz_mat_eq}) as $[A_{1},A_{2},...,A_{K}]$, where $K=1,2,4$
or $8$ and each $A_{k}$ is a $K\times K$ matrix. Suppose $N=\text{dim}(\boldsymbol{a})$
satisfies $N=qK$ for a positive integer $q$. We define a sparse
$K$-compositional binding by
\begin{equation}
P_{i}=\underbrace{O_{K}\oplus...\oplus O_{K}}_{\left\lfloor (i-1)/K\right\rfloor }\oplus A_{i\%K}\oplus O_{K}\oplus...\oplus O_{K},\label{eq_def_sparse_Kcomp}
\end{equation}
for $i=1,...,N$, where $O_{K}$ is the $K\times K$ zero matrix,
$A\oplus B$ is the direct sum of matrices $A$ and $B$, $\left\lfloor x\right\rfloor $
represents the largest integer smaller or equal to $x$, and $x\%y$
is the remainder of $x$ divided by $y$ (for ease of notation, we
define $A_{0}=A_{K}$). We denote this binding mechanism as the sparse
$K$-compositional binding because it is a sparse implementation of
the composition algebra of dimension $K$. For instance, if $K=2$
and $q=3$, we get 
\begin{equation}
P_{1}=\left(\begin{array}{ccc}
A_{1} & O & O\\
O & O & O\\
O & O & O
\end{array}\right),\;P_{2}=\left(\begin{array}{ccc}
A_{2} & O & O\\
O & O & O\\
O & O & O
\end{array}\right),\;P_{3}=\left(\begin{array}{ccc}
O & O & O\\
O & A_{1} & O\\
O & O & O
\end{array}\right),\;P_{4}=\left(\begin{array}{ccc}
O & O & O\\
O & A_{2} & O\\
O & O & O
\end{array}\right),...
\end{equation}
When $q=1$, they are matrix representations of the composition algebra
we discussed in the previous section. Whereas under $K=1$, we get
a binding by the Hadamard product. Although $P=\left[P_{1},P_{2},...,P_{N}\right]$
does not satisfy the Hurwitz matrix equations when $q>1$ ($P_{i}P_{i}^{T}\neq\lambda I_{N}$),
it satisfies the fixed-point condition, Eqs. \ref{eq_dla_dp_sym}
and \ref{eq_dlb_dp_sym} (see Appendix \ref{subsec:sparse_Kcomp}).
This means that a sparse $K$-compositional tensor $P$ is a fixed-point
solution of both $\ell_{a}$ and $\ell_{b}$. The decoding error under
this binding is given as
\begin{equation}
\ell_{a}=\ell_{b}=\frac{2}{K+2}.
\end{equation}
This error is significantly smaller than that of HRR under $K=4$
and $K=8$ (Fig. 4A, purple and orange lines vs. blue line). In particular,
under $K=8$, we get $\ell_{a}=\ell_{b}=1/5$, the same error we observed
under a numerical optimization (compare Figs. 2A and B with Fig. 4A).
It also explains why we found $8\times8$ block structures (Figs.
3C and D). Among sparse $K$-compositional bindings, $K=8$ yields
the smallest error because it is the largest matrix family that satisfies
the Hurwitz matrix equations.

Using Cayley-Dickson construction \citep{baez2002octonions}, we can
in principle construct sparse $K$-compositional binding tensors for
$K=16$ (sedenions), $32$ (trigintaduonions), and so on. However,
it does not improve the unbinding performance (right side of Fig.
4B), because they do not satisfy the Hurwitz matrix equations. Thus,
the binding tensor with $K=8$ provides the best unbinding performance
among the sparse $K$-compositional bindings. Below, we denote this
$K=8$ solution of Eq. \ref{eq_def_sparse_Kcomp} as the octonion
binding, because it employs a matrix representation of the octonions.

The octonion binding solution we constructed is not the same with
the numerically optimized one, in a sense that only one of the block
diagonal components is non-zero, while all block diagonal components
are non-zero in the numerically optimized matrices (Figs. 3C and D).
However, we found that the error of the sparse $K$-compositional
binding is conserved under a transformation with any orthogonal matrix,
which makes all block diagonal components non-zero (Appendix \ref{subsec:sparse_Kcomp}).
This result suggests that the numerically optimized solutions are
consistent with the octonion binding. Note that, unlike $K$-compositional
bindings with $K\leq8$, HRR does not satisfy Eqs. \ref{eq_dla_dp_sym}
and \ref{eq_dlb_dp_sym} at a finite $N$ (see Appendix \ref{subsec:HRR}). 

\begin{figure}
\begin{centering}
\includegraphics[scale=0.9]{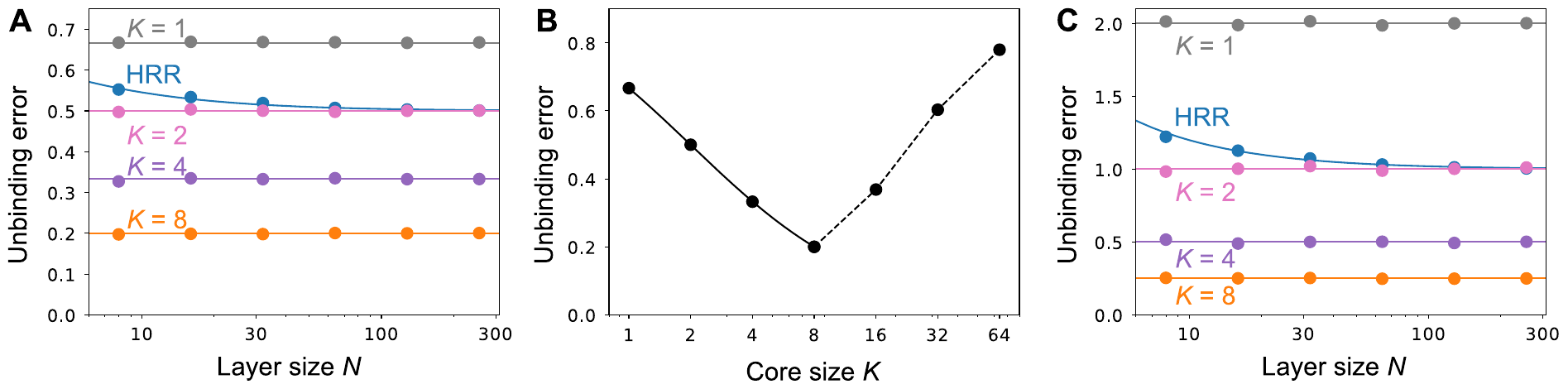}\caption{Performance of sparse $K$-compositional binding methods under $L=1$.
\textbf{A)} Decoding error $\ell_{a}$ of sparse $K$-compositional binding methods with the core size $K=1,2,4,8$ and the holographic reduced representation (HRR) under various layer sizes $N$. Points are simulations and lines are theory ($\ell_{a}=2/(K+2)$ for $K$-compositional, and $\ell_{a}=\frac{N+2}{2(N+1)}$ for HRR). 
\textbf{B)} Decoding error $\ell_{a}$ of the sparse $K$-compositional binding methods with various core sizes $K$ under $N=128$. The solid line is $\frac{2}{K+2}$, while the dotted line is a linear interpolation. 
\textbf{C)} The same as \textbf{A}, except that the binding operators were normalized so that the amplitude of the signal is maintained in the unbinding. Points are simulations and lines are theory ($\ell_{a}=2/K$ for $K$-compositional, and $\ell_{a}=1+\frac{2}{N}$ for HRR). }
\rule{\textwidth}{0.25pt}
\par\end{centering}
\end{figure}

A binding operator $P$ derived by minimizing $\ell_{a}$ shrinks
the amplitude of the signal in the reconstructed vector $\hat{\boldsymbol{a}}_{1}$
as shown in Eq. \ref{eq_quarternion_unbinding} ($\left\langle \hat{\boldsymbol{a}}\right\rangle _{p(b)}=\frac{2}{3}\boldsymbol{a}$
under the quaternion binding), meaning that the decoding is biased.
However, it might be desirable to use an unbiased binding operator,
which keeps the signal amplitude in the decoded vector $\hat{\boldsymbol{a}}_{1}$
the same with the original signal $\boldsymbol{a}_{1}$ (i.e., $\left\langle \hat{\boldsymbol{a}}_{1}-\boldsymbol{a}_{1}\right\rangle =0$).
Taking the expectation over $\boldsymbol{b}$, the reconstructed vector
$\hat{a}_{i}^{1}$ (Eq. \ref{eq_def_qunbind_a}) on average satisfies
\begin{equation}
\left\langle \hat{a}_{i}^{1}\right\rangle _{p(b)}=\sum_{j=1}^{N}\sum_{l=1}^{N}\left(\sum_{k=1}^{N}Q_{ijk}P_{ljk}\right)a_{l}^{1}.
\end{equation}
Thus, in order to keep the reconstructed signal amplitude the same
with the original signal, under $P=Q$, $P$ should be normalized
as
\begin{equation}
\sum_{j=1}^{N}\sum_{k=1}^{N}P_{ijk}^{2}=1,
\end{equation}
for $i=1,2,...,N$. Under this normalization, the unbinding error
tends to be larger, but the relative performance of various binding
methods is preserved (Fig. 4C vs. Fig. 4A). In particular, the sparse
$K$-compositional bindings with $K=4$ and $8$ still outperform
HRR with the same unbiased normalization (orange and purple lines
vs. blue line in Fig. 4C). 

To gain further insights into the space of the quadratic binding mechanisms,
we next study the stability of the quaternion and the octonion bindings
(Eq. \ref{eq_def_sparse_Kcomp} with $K=4$ and $8$, respectively)
against a perturbation. If we initialize $P,Q,R$ as the quaternion
binding plus a small random perturbation, then run the iterative optimization
process (Algorithm 1), the error converges to $1/3$, the original
error level under the quaternion binding (blue lines in Fig. 5A).
On the other hand, under a large perturbation, the error instead converges
to $1/5$, the error level under the octonion binding (purple and
pink lines). These results indicate that the quaternion binding is
a local minimum in the space of binding operators. The octonion binding
is, on the contrary, stable against perturbation (Fig. 5B), although
$P,Q,R$ converge to different tensors when a large perturbation is
added. This result suggests that the octonion binding has a large
basin of attraction in the parameter space, though it only indicates
a local optimality of the octonion binding, not a global one. 

\begin{figure}
\begin{centering}
\includegraphics[scale=0.9]{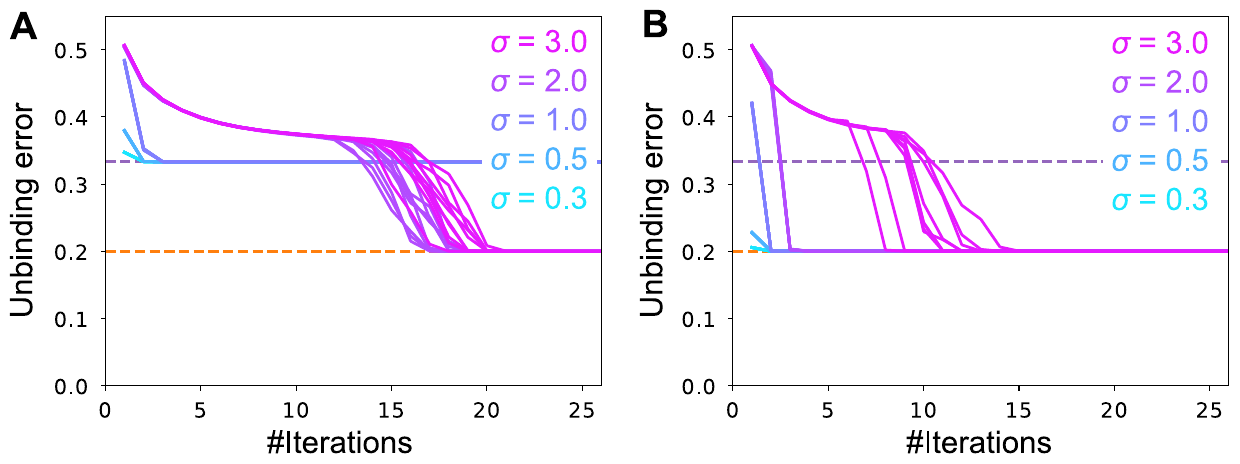}\caption{Local stability of the quaternion and the octonion bindings at $N=48$ and $L=1$. 
\textbf{A)} Learning curve from the quaternion binding ($K=4$) plus perturbation. We constructed the initial $P,Q,R$ by adding random Gaussian noise with the standard deviation $\sigma/N$ to the sparse-quaternions binding. 
\textbf{B)} Learning curve from the octonion binding ($K=8$) plus
perturbation. Both in \textbf{A} and \textbf{B}, we measure the unbinding error by $\ell_{a}$.}
\rule{\textwidth}{0.25pt}
\par\end{centering}
\end{figure}

\section{The binding solutions under $L>1$ }

Our theoretical and numerical analyses in the previous section suggest
that a binding method based on a matrix representation of octonions,
the octonion binding, outperforms HRR binding under $L=1$. How does
this method scale to an unbinding from a composition $\boldsymbol{c}$
that consists of multiple bound pairs ($L>1$)? 

The numerical optimization method (Algorithm 1) can be straightforwardly
applied to $L>1$. For instance, an update of $P$ with respect to
loss $\ell_{a}$ is done by $\text{Vec}\left[P_{l}\right]=\left(\Gamma^{q,L}\right)^{-1}\text{Vec}\left[Q_{l}\right]$,
where $N^{2}\times N^{2}$ matrix $\Gamma^{q,L}$ is defined as (see
Appendix \ref{subsec:numerical_opt}),
\begin{equation}
\Gamma_{[jN+k],[mN+n]}^{q,L}\equiv\sum_{i}\left(\delta_{jm}L\sum_{\beta}Q_{i\beta k}Q_{i\beta n}+Q_{ijk}Q_{imn}+Q_{ijn}Q_{imk}\right).
\end{equation}
Applying this numerical optimization to the case when $L=3$, we found
$8\times8$ block diagonal structures in the converged binding matrices
as before (Figs. 6B and C). The decoding performance
of the obtained binding is better than HRR (black vs. gray dashed
lines in Fig. 6A), though the relative advantage was
smaller compared to the case when $L=1$ (Fig. 6A
vs. Fig. 2A). Moreover, the octonion binding satisfies the fixed-point
condition for both $\ell_{a}$ and $\ell_{b}$ even when $L>1$ (see
Appendix \ref{subsec:sparse_Kcomp}). These results indicate that
the octonion binding may have an edge even when $L>1$. However, it
is only suggestive because both the octonion and HRR bindings perform
poorly when we directly apply them to $L>1$ (Fig. 6A;
noise-to-signal ratio is above 0.7 for both). 
\begin{figure}
\begin{centering}
\includegraphics[scale=0.9]{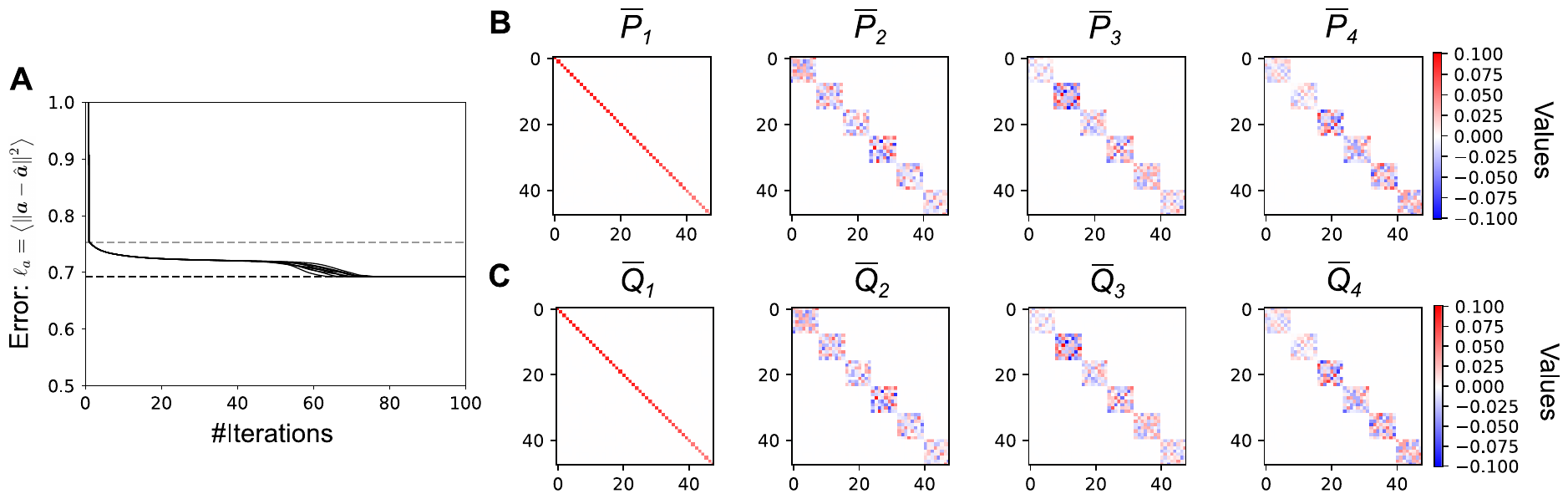}\caption{\label{fig_opt_L3}Numerical optimization of $P,Q,R$ under $N=48$,
$L=3$. \textbf{A)} Learning dynamics from ten random initializations (the same as Fig. 2A, but under $L=3$). Black and gray horizontal dashed lines are the error under the octonion binding ($\ell=\frac{9}{13}$), and HRR ($\ell=\frac{3}{4}$), respectively. 
\textbf{B, C)} Examples of the learned binding matrices after 300 iterations. The panels are the same with Figs. 3C and 3D, but calculated under $L=3$. Note that $\bar{P}_{i}=\bar{Q}_{i}$ holds for all $i$. }
\rule{\textwidth}{0.25pt}
\par\end{centering}
\end{figure}

Below, we first show that the performance of all unbiased quadratic
binding operators is lower bounded by $\ell_{a}\geq\frac{NL}{N_{c}}+\mathcal{O}\left(1\right)$
under a mild condition, hence equi-sized binding operators (i.e.,
$N_{c}=N$), such as HRR or the octonion binding, inevitably scale
poorly under $L>1$. To overcome this problem, we consider two extensions:
decoding with a dictionary \citep{plate1995holographic,smolensky2014optimization}
and decoding from a composition $\boldsymbol{c}$ larger than the
size of elements $\boldsymbol{a}$ and $\boldsymbol{b}$ ($N_{c}>N$)
\citep{smolensky1990tensor,frady2020variable}. We show that under
both extensions, the proposed octonion binding outperforms both HRR
and a random binding under $L\sim\mathcal{O}\left(1\right)$, but
its advantage disappears at the large $L$ limit. 

\subsection{Lower bound on the unbinding error }

Let us first focus on the case when $L\gg1$, and consider minimization
of the decoding error $\ell_{a}$ under a general quadratic parameterization.
As we discussed previously, in order to retain the signal amplitude
in the unbinding process, and thus to make the estimation unbiased,
from Eq. \ref{eq_unbinding_amplitude}, $P$ and $Q$ need to be normalized
as 
\begin{equation}
\sum_{j=1}^{N}\sum_{k=1}^{N_{c}}P_{ijk}Q_{ijk}=1.
\end{equation}
Taking the large $L$ limit of the loss $\ell_{a}$ under this constraint,
as a function of $L$, the loss $\ell_{a}$ follows (see Appendix
\ref{subsec:Lower-bound})
\begin{equation}
\frac{\ell_{a}}{L}=\frac{1}{N}\sum_{i=1}^{N}\sum_{l=1}^{N}tr\left[P_{l}Q_{i}^{T}Q_{i}P_{l}^{T}\right]+\mathcal{O}\left(\frac{1}{L}\right).
\end{equation}
Thus, at $L\gg1$, minimization of the loss $\ell_{a}$ under the
signal amplitude constraint is reformulated as the minimization of
a Lagrangian:
\begin{equation}
\mathscr{\mathcal{L}}_{a}=\frac{1}{2}\sum_{i=1}^{N}\sum_{l=1}^{N}tr\left[P_{l}Q_{i}^{T}Q_{i}P_{l}^{T}\right]-\sum_{i=1}^{N}\lambda_{i}\left(tr\left[P_{i}Q_{i}^{T}\right]-1\right).
\end{equation}
Solving this Lagrangian under an assumption that $\sum_{i}P_{i}P_{i}^{T}$
is invertible, and applying Jensen's inequality, the lower bound of
$\ell_{a}$ is given as 
\begin{equation}
\frac{\ell_{a}}{L}\geq\frac{N}{N_{c}}+\mathcal{O}\left(\frac{1}{L}\right),\label{eq_la_bound}
\end{equation}
where, recall that, $N\equiv\dim(\boldsymbol{a})=\dim(\boldsymbol{b})$,
and $N_{c}\equiv\dim(\boldsymbol{c})$. This means that the error
under any quadratic bindings satisfying the invertibility condition
is lower bounded by $\ell_{a}\geq\frac{NL}{N_{c}}+C$, where $C$
is a term that does not depend on $L$. In particular, under an equi-sized
composition ($N_{c}=N$), the error is bounded by $\ell_{a}\geq L+C$. 

Figure 7 describes the unbinding error $\ell_{a}$
as a function of the number of bound pairs $L$ under $N_{c}=N$ for
three different binding mechanisms. Under the octonion binding, the
error follows $\ell_{a}=L-\frac{3}{4}$ (orange line in Fig. 7),
whereas under HRR, $\ell_{a}=L+\frac{2}{N}$ (blue line). It means
that, both binding methods tightly follow the lower bound (Eq. \ref{eq_la_bound})
though the octonion binding has a smaller intercept $C$ than HRR.
On the contrary, the sum binding ($\boldsymbol{c}=\sum_{\mu}\left(\boldsymbol{a}_{\mu}+\boldsymbol{b}_{\mu}\right)$
and $\hat{\boldsymbol{a}}_{1}=\boldsymbol{c}-\boldsymbol{b}_{1}$)
yields $\ell_{a}=2L-1$ (green line in Fig. 7). Thus,
it performs progressively worse compared to the two other methods
as $L$ becomes larger, though it has the smallest error under $L=1$.
Notably, all three methods yield errors larger than one for $L>1$,
meaning that the signal-to-noise ratio is smaller than one. Therefore,
we need to modify these methods to perform decoding from a composition
of multiple pairs. Below, we first consider decoding with a help of
a dictionary, then study binding with an expanded composition ($N_{c}>N$). 

\begin{figure}
\centering{}\includegraphics[scale=0.9]{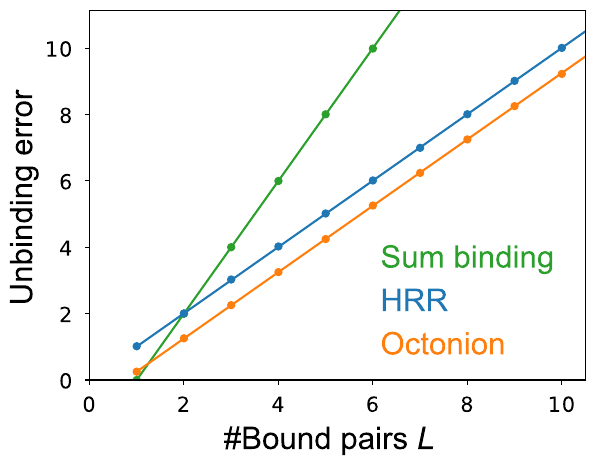}\caption{\label{fig_L_scale}Comparison of the loss $\ell_{a}$ under the octonion,
HRR, and the sum bindings at $N_{c}=N=128$ for various $L$. Points
are simulations and lines are theory ($\ell_{a}=2L-1$, $L+\frac{2}{N}$,
and $L-\frac{3}{4}$ for the sum-binding, HRR, and the octonion, respectively). }
\rule{\textwidth}{0.25pt}
\end{figure}

\subsection{Decoding with a dictionary }

Previous studies showed that if the system knows the dictionary from
which vectors $\left\{ \boldsymbol{a}_{\mu}\right\} _{\mu=1}^{L}$
and $\left\{ \boldsymbol{b}_{\mu}\right\} _{\mu=1}^{L}$ are sampled,
accurate decoding is possible even if multiple pairs are bound together
\citep{plate1995holographic,smolensky2014optimization}. Hence, we
introduce a dictionary containing $D$ words $\mathcal{D}=\left\{ \boldsymbol{a}_{d}\right\} _{d=1}^{D}$,
where each word $\boldsymbol{a}_{d}$ is sampled from an i.i.d. Gaussian
distribution $\mathcal{N}\left(0,I_{N}\right)$. With this dictionary,
we conduct unbinding of $\boldsymbol{a}_{1}$ from a composition $\boldsymbol{c}$
with a query $\boldsymbol{b}_{1}$ in the following steps:
\begin{enumerate}
\item Unbind $\boldsymbol{a}_{1}$ by $\hat{a}_{i}^{1}=\sum_{j}\sum_{k}Q_{ijk}b_{j}^{1}c_{k}$,
as before.
\item Calculate $z_{\mu}=\hat{\boldsymbol{a}}\cdot\boldsymbol{a}_{\mu}$
for all the words in the dictionary ($\mu=1,...,D$).
\item Pick the word $\boldsymbol{a}_{\mu_{o}}$ with $\mu_{o}\equiv\textrm{argmax}_{\mu}z_{\mu}$.
\end{enumerate}
Note that, because all $a_{\mu}$ have roughly the same norm ($\left\Vert \boldsymbol{a}_{\mu}\right\Vert ^{2}\approx N$),
$\textrm{argmax}_{\mu}\hat{\boldsymbol{a}}\cdot\boldsymbol{a}_{\mu}$
is equivalent to $\textrm{argmin}_{\mu}\left\Vert \hat{\boldsymbol{a}}-\boldsymbol{a}_{\mu}\right\Vert ^{2}$
at the large $N$ limit. Approximating $p\left(z_{1},...,z_{D}\right)$
with a Gaussian distribution, the probability of a misclassification
is given as (see Appendix \ref{subsec:dictionary_method})
\begin{equation}
P\left[\mu_{o}\neq1\right]=1-\int\frac{dz}{\sqrt{2\pi}}\exp\left[-\frac{1}{2}\left(z-\frac{1}{\sigma_{s}}\right)^{2}\right]\left(\Phi\left[\frac{\sigma_{s}z}{\sigma_{l}}\right]\right)^{L-1}\left(\Phi\left[\frac{\sigma_{s}z}{\sigma_{d}}\right]\right)^{D-L},
\end{equation}
where $\Phi\left[z\right]$ is the cumulative Gaussian distribution,
and $\sigma_{s}$, $\sigma_{l}$, $\sigma_{d}$ are the standard deviations
of $z_{\mu}$ under $\mu=1$ (the target), $2\leq\mu\leq L$ (the
bound words), and $L+1\leq\mu\leq D$ (the rest of words in the dictionary),
respectively. 

The standard deviations $\sigma_{s}$, $\sigma_{l}$ and $\sigma_{d}$
depend on the choice of the binding and unbinding methods. For instance,
under the octonion binding, we get 
\begin{equation}
\sigma_{s}^{2}=\frac{1}{N}\left(L+\frac{3}{2}\right),\quad\sigma_{l}^{2}=\frac{1}{N}\left(L+\frac{1}{2}\right),\quad\sigma_{d}^{2}=\frac{1}{N}\left(L+\frac{1}{4}\right),\label{eq_var_dict_octonion}
\end{equation}
whereas under HRR, assuming $N\gg1$, 
\begin{equation}
\sigma_{s}^{2}=\frac{1}{N}\left(L+3\right),\quad\sigma_{l}^{2}=\frac{1}{N}\left(L+2\right),\quad\sigma_{d}^{2}=\frac{1}{N}\left(L+1\right).\label{eq_var_dict_hrr}
\end{equation}
Therefore, we expect the classification error of two binding methods
to be different under $L\sim\mathcal{O}\left(1\right)$, though the
performance should converge to the same accuracy under $L\gg1$. In
particular, a small $\sigma_{s}$ effectively amplifies the signal
$N/\sigma_{s}$, while a small $\sigma_{s}/\sigma_{d}$ makes a misclassification
with non-bound words in the dictionary less likely (note that $\left[\sigma_{s}/\sigma_{d}\right]_{oct}\leq\left[\sigma_{s}/\sigma_{d}\right]_{HRR}$
for $L\geq1$). 

Figure 8A describes the probability of incorrect classification
($P\left[\mu_{o}\neq1\right]$) under the octonion binding and HRR,
and also under a random binding introduced as a control. The random
binding was constructed by sampling the elements of $P$ from a Gaussian
distribution $\mathcal{N}\left(0,1/N^{2}\right)$ and setting $Q=R=P$.
As in the case of unbinding without a dictionary, the performance
becomes worse as the number of bound pairs goes up (compare Fig. 8A
with Fig. 7). However, even when a dozen pairs are
bound to the composition, the misclassification rate is far below
the chance level ($P\left(\mu_{o}\neq1\right)=1-\frac{1}{D}$) in
all three binding methods. Moreover, we found that the error under
the octonion binding is smaller than a random binding under a small
$L$ (orange vs. red in Fig. 8A), while that of HRR
is roughly the same with the random binding (blue vs. red; some blue
points are hidden under the red points). 

In the comparison above, all three methods perform worse when a large
number of pairs are bound to the composition, making the comparison
difficult in this regime. To clarify the issue, using the fact that
the signal-to-noise ratio roughly scales with $N/L$ (Eqs. \ref{eq_var_dict_octonion}
and \ref{eq_var_dict_hrr}), we plotted the misclassification probability
for different $L$ while scaling the vector size as $N\propto L$
(Fig. 8B). In this parameterization, the octonion binding
exhibits similar error curves as a function of the dictionary size
$D$ regardless of $L$ (orange lines). On the contrary, HRR and the
random binding show higher errors under $L=5$ and $L=10$ (orange
vs. blue and red points in Fig. 8B left and middle panels),
though their performance converges to that of the octonion binding
under $L=20$ (right panel). This result suggests that the performance
of HRR and the random binding are suboptimal under $L\sim\mathcal{O}\left(1\right)$,
while all three methods perform similarly under a large $L$.

Because the number of things humans can keep in the working memory
is suggested to be less than ten \citep{miller1956magical,cowan2001magical},
in a cognitive process that requires binding in the working memory
such as scene understanding, we expect $L\leq10$ to be the biologically
relevant parameter regime. Our result indicates that the octonion
binding outperforms HRR and the random binding in this regime, though
its advantage goes away under a large $L$. 

\begin{figure}
\centering{}\includegraphics[scale=0.9]{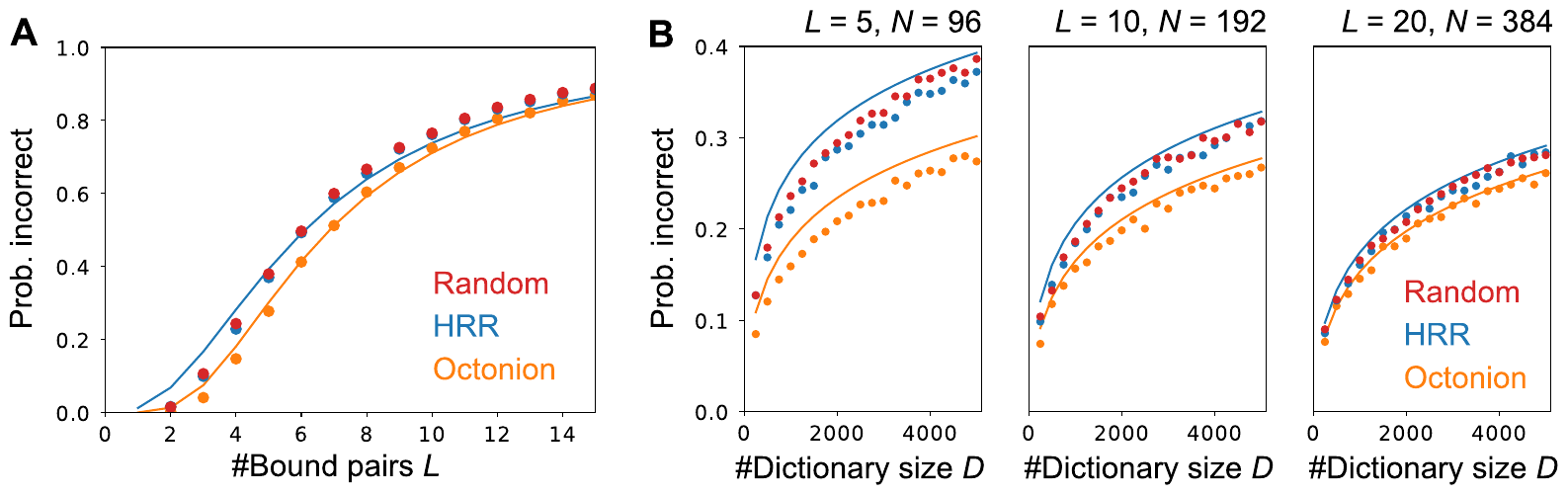}\caption{\label{fig_dict}Decoding performance in the presence of a dictionary.
\textbf{A)} Error rates (probability of incorrect classification) of the octonion, HRR and the random bindings, under various various number of bound pairs $L$. 
We set the number of the items in the dictionary to $D=5000$ and the vector size as $N_{c}=N=96$. Points are simulations and lines are theory (see Appendix \ref{subsec:dictionary_method} for the details).
Analytical lines for the random binding is omitted because that is exactly the same as the line for HRR. 
\textbf{B)} Error rates of the octonion, HRR and the random bindings, under various dictionary sizes $D$.
Three panels represent the errors under different $L$ and $N$ under
a fixed ratio $N/L$. Points are simulations and lines are theory.
The lines deviate from the points under $L=5$ because the Gaussianity assumption made for the theoretical lines is violated in this regime.}
\rule{\textwidth}{0.25pt}
\end{figure}

\subsection{Extension of the sparse octonion binding and HRR to $N_{c}>N$}

Even if the brain does not know the dictionary from which words are
sampled, it can achieve a good decoding performance with an expansion
of the composition layer. Indeed, Eq. \ref{eq_la_bound} indicates
that, if the size of composition $N_{c}$ scales with the (maximum)
number of bound pairs $L$, the system can reliably perform an unbinding
from a composition of multiple pairs. Thus, we next consider an extension
of the octonion, HRR, and random binding mechanisms to $N_{c}>N$. 

When $N_{c}$ is a multiple of $N$ but smaller than $N^{2}/8$, the
octonion binding is straightforwardly extended to $N_{c}>N$ by adding
shifted block-diagonal components (see Appendix \ref{subsec:Extended-Octonion-binding}).
Under this extended octonion binding mechanism, the decoding error
becomes $\ell_{a}=\frac{N}{N_{c}}\left(L-1+\frac{2}{K}\right)$ (orange
lines in Figs. 9A and B). Notably, the leading term of
$\ell_{a}$ with respect to $L$ is still the same with the lower
bound (Eq. \ref{eq_la_bound}). 

HRR can also be extended to $N_{c}>N$ by considering an interpolation
of HRR and the tensor product representation \citep{smolensky1990tensor}.
Because the binding tensor $P$ is given as $P_{ijk}=\frac{1}{\sqrt{N}}\delta_{[i+j]_{N},k}$
for HRR and $P_{ijk}=\frac{1}{\sqrt{N}}\delta_{(iN+j),k}$ for the
tensor product representation, we can interpolate these two bindings
by setting $P_{ijk}=\frac{1}{\sqrt{N}}\delta_{[id+j]_{dN},k}$ for
$N_{c}=dN$ (see Appendix \ref{subsec:Tensor-HRR-morphing}). The
decoding error approximately follows $\ell_{a}\approx\frac{NL}{N_{c}}+\frac{1}{N}$
under this tensor-HRR binding (blue lines in Figs. 9A
and B, partially occluded by the red lines). 

Lastly, an extension of the random binding is done straightforwardly
by sampling the elements of $P$ from a Gaussian distribution with
the mean zero and the variance $1/(NN_{c})$, while setting $Q=R=P$.
Under this method, assuming $N,N_{c}\gg1$, the decoding error becomes
(see Appendix \ref{subsec:random_binding}), 
\begin{equation}
\ell_{a}=\frac{LN}{N_{c}}\left(1+\frac{N_{c}}{N^{2}}\right).\label{eq_la_random2}
\end{equation}
Thus, at $N_{c}\ll N^{2}$, the leading order term of the error follows
the lower bound $LN/N_{c}$ (red lines in Figs. 9A and
B) while at $N_{c}\to N^{2}$ limit, the error of this
random binding becomes the double of the lower bound. 

Because it has a small intercept, the extended octonion binding outperforms
both tensor-HRR interpolation and the random binding under $L\sim\mathcal{O}\left(1\right)$
(Fig. 9A; here $L=3$). This result again indicates that
the octonion-based binding method is preferable in the parameter regime
relevant to working memory-based cognitive processes. However, at
$L\gg1$, the random binding is as good as the extended octonion bindings
(Fig. 9B; $L=10$). In fact, comparing the lower bound
(Eq. \ref{eq_la_bound}) with Eq. \ref{eq_la_random2}, we can conclude
that there is no quadratic binding method with invertible $\sum_{i}P_{i}P_{i}^{T}$
that significantly outperforms the random binding when $L\gg1$ and
$N_{c}\ll N^{2}$. 

\begin{figure}
\begin{centering}
\includegraphics[scale=0.9]{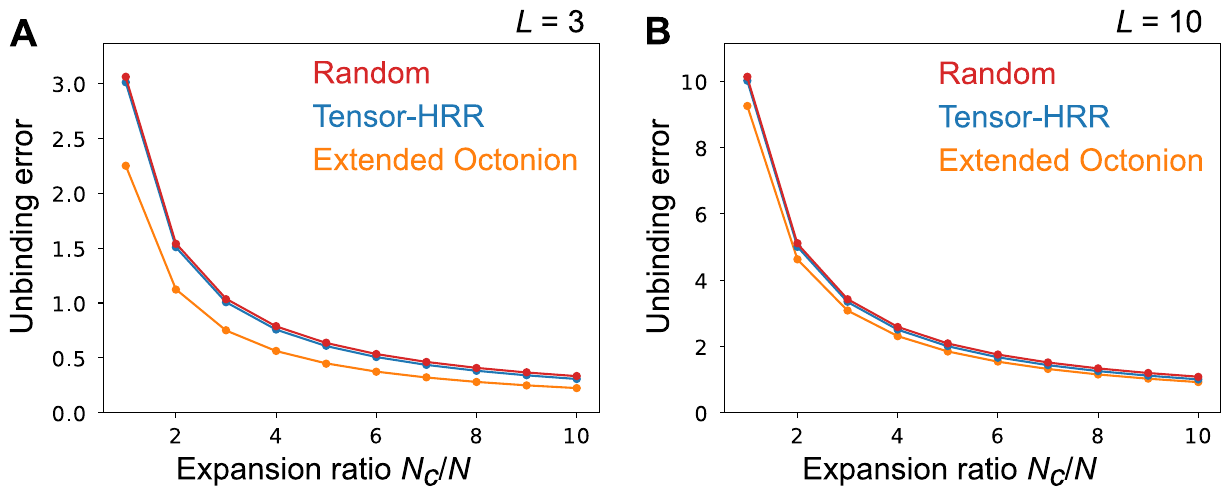}
\par\end{centering}
\caption{\label{fig_ext}Comparison of the random, tensor-HRR, and the extended octonion bindings under various expansion ratios $N_{c}/N$ at $L=3$ (\textbf{A}) and $L=10$ (\textbf{B}). 
We set $N=128$. Points are simulation results,
and lines are theoretical results from Eqs. \ref{eq_la_extended_octonions} (extended octonions), \ref{eq_la_tenHRR} (tensor-HRR), and \ref{eq_la_random} (random).}
\rule{\textwidth}{0.25pt}
\end{figure}

\section{Discussion}

In this work, we investigated optimal methods for pair-wise binding
based on the VSA framework \citep{smolensky1990tensor,plate1995holographic,gayler2004vector}.
We first numerically optimized the binding and unbinding operators
for the best unbinding performance assuming only one pair is bound
to the composition vector. We found that the numerically optimized
binding operators outperform HRR, a popular method for binding (Fig.
2). Moreover, we revealed that there is a hidden $8\times8$ block
structure in the optimized binding and unbinding matrices (Fig. 3).
By analytically deriving a sufficient condition for a fixed-point
of the loss function, we show that the $8\times8$ block structure
is originated from a matrix representation of the octonion algebra,
an eight-dimensional extension of the complex numbers (Fig. 4). Furthermore,
we showed that even when several pairs are bound into a composition,
the proposed binding method based on the octonion outperforms previously
proposed methods both under the dictionary decoding (Fig. 8)
and unbinding from an expanded composition (Fig. 9).
When there are many bound pairs in a composition, however, the advantage
of the proposed method vanishes, and even a random binding shows approximately
the optimal unbinding performance under a mild condition (Figs. 8
and 9). 

We introduced two key assumptions for deriving these conclusions:
Both binding and unbinding operators have quadratic forms, and input
vectors are i.i.d. random Gaussian vectors. The former assumption
is reasonable under the latter assumption because a quadratic binding
should be enough to capture the statistical relationship between the
inputs when the inputs are Gaussian. We leave an investigation of
the optimal binding under general input statistics for future works.

Many binding mechanisms have been proposed previously in the framework
of VSA \citep{smolensky1990tensor,plate1995holographic,kanerva1997fully,gallant2013representing,gosmann2019vector,frady2020variable}.
In particular, Frady and colleagues proposed a block-wise circular
convolution method to conserve the sparsity of the composition \citep{frady2020variable}.
At the limit where each block is $2\times2$ matrix, their binding
method corresponds to our sparse $K$-compositional binding with $K=2$
in which a matrix representation of the complex numbers is used for
binding. However, their analysis is limited to the case when the input
is a block-wise one-hot vector, and they did not investigate other
block-wise binding mechanisms. In addition, the relationship between
the Clifford algebra, a generalization of the quaternion algebra,
and HRR was previously investigated by Aerts and colleagues \citep{aerts2009geometric}.
However, they did not study the space of binding mechanisms or the
optimization of binding methods. 

Recent experimental results found a mixed representation of sensory
stimuli and context cues in the prefrontal cortex \citep{rigotti2013importance}
and hippocampus \citep{nieh2021geometry}. However, it remains unclear
whether mixed representation in the brain is random or structured
\citep{hirokawa2019frontal}. Our results suggest that depending on
the task configuration, random binding might be enough for an accurate
unbinding, though it is unclear if unbinding is crucial in the tasks
employed in these experiments. 

The binding problem is also an important topic in machine learning
\citep{greff2020binding}. In knowledge graph embedding tasks \citep{nickel2015review},
Nickel and colleagues showed that HRR yields a better generalization
performance than methods based on nonlinear-projection of a concatenated
vector \citep{nickel2016holographic}. Moreover, in visual question
answering tasks, binding of an image representation and a query representation
is crucial for solving the task. The Hadamard product is often employed
for this binding \citep{antol2015vqa,santoro2017simple}, but more
elaborate binding mechanisms, such as self-attention on concatenated
vectors, are suggested to improve the learning performance \citep{teney2018tips}.

Lastly, unlike the quaternions, the octonions are rarely applied to
the domain of science \citep{baez2002octonions}. Our work provides
a rare practical application of octonion algebra. More generally,
our work indicates a potential link between the mathematics of quadratic
forms and the binding problem in cognitive science and machine learning. 

\appendix

\section*{{\LARGE Appendix}}
\section{Quadratic binding}

\subsection{Fixed-point condition of the mean-squared error\label{subsec:Fixed-point-condition}}

From Eqs. \ref{eq_def_qbind}, \ref{eq_def_qunbind_a}, and \ref{eq_def_la},
the loss function $\ell_{a}$ is written as 
\begin{equation}
\begin{aligned}\ell_{a} & =\frac{1}{N}\left\langle \sum_{i=1}^{N}\left(a_{i}^{1}-\hat{a}_{i}^{1}\right)^{2}\right\rangle _{p(S)}\\
 & =\frac{1}{N}\left\langle \sum_{i=1}^{N}\left(a_{i}^{1}-\sum_{\mu=1}^{L}\sum_{j=1}^{N}\sum_{l=1}^{N}\sum_{m=1}^{N}\left[\sum_{k=1}^{N_{c}}Q_{ijk}P_{lmk}\right]b_{j}^{1}b_{m}^{\mu}a_{l}^{\mu}\right)^{2}\right\rangle _{p(S)}
\end{aligned}
\end{equation}
where $\left\langle \;\cdot\;\right\rangle _{p(S)}$ is the expectation
over random vectors $\boldsymbol{a}_{\mu}$ and $\boldsymbol{b}_{\mu}$
sampled i.i.d. from a Gaussian distribution $N(\boldsymbol{0},I_{N})$.
For simplicity, let us introduce a fourth-order tensor $M$ as, 
\begin{equation}
M_{mj}^{li}\equiv\sum_{k=1}^{N_{c}}Q_{ijk}P_{lmk}.\label{eq_def_tensorM}
\end{equation}
Then the loss is rewritten as
\begin{equation}
\begin{aligned}\ell_{a} & =\frac{1}{N}\left\langle \sum_{i=1}^{N}\left(\left(a_{i}^{1}\right)^{2}-2\sum_{\mu=1}^{L}\sum_{j=1}^{N}\sum_{l=1}^{N}\sum_{m=1}^{N}M_{mj}^{li}a_{i}^{1}a_{l}^{\mu}b_{j}^{1}b_{m}^{\mu}+\left(\sum_{\mu=1}^{L}\sum_{j=1}^{N}\sum_{l=1}^{N}\sum_{m=1}^{N}M_{mj}^{li}a_{l}^{\mu}b_{j}^{1}b_{m}^{\mu}\right)^{2}\right)\right\rangle _{p(S)}\end{aligned}
\end{equation}
The expectation over the last quadratic term becomes 
\begin{equation}
\begin{aligned} & \left\langle \sum_{i=1}^{N}\left(\sum_{\mu=1}^{L}\sum_{j=1}^{N}\sum_{l=1}^{N}\sum_{m=1}^{N}M_{mj}^{li}a_{l}^{\mu}b_{j}^{1}b_{m}^{\mu}\right)^{2}\right\rangle \\
 & =\sum_{\mu=2}^{L}\sum_{i}\sum_{l}\sum_{j}\sum_{m}\left(M_{mj}^{li}\right)^{2}+\sum_{i}\sum_{l}\left(\sum_{j}\sum_{m}M_{mj}^{li}M_{mj}^{li}+\sum_{j}\sum_{m}M_{mj}^{li}M_{jm}^{li}+\sum_{j}\sum_{m}M_{jj}^{li}M_{mm}^{li}\right)\\
 & =\sum_{i}\sum_{l}\left(\left(\sum_{j}M_{jj}^{li}\right)^{2}+\sum_{j}\sum_{m}M_{mj}^{li}\left[L\cdot M_{mj}^{li}+M_{jm}^{li}\right]\right),
\end{aligned}
\end{equation}
where summation runs from $1$ to $N$ unless otherwise stated. In
the second line, we used
\begin{equation}
\left\langle b_{i}^{1}b_{j}^{1}b_{k}^{1}b_{l}^{1}\right\rangle =\delta_{ij}\delta_{kl}+\delta_{ik}\delta_{jl}+\delta_{il}\delta_{jk}.
\end{equation}
Thus, the loss $\ell_{a}$ is written as
\begin{equation}
\ell_{a}=1-\frac{2}{N}\sum_{i}\sum_{j}M_{jj}^{ii}+\frac{1}{N}\sum_{i}\sum_{l}\left(\left({\textstyle \sum_{j}}M_{jj}^{li}\right)^{2}+\sum_{j}\sum_{m}M_{mj}^{li}\left[L\cdot M_{mj}^{li}+M_{jm}^{li}\right]\right),\label{eq_la_full}
\end{equation}
Using $P$ and $Q$ instead of $M$ via Eq. \ref{eq_def_tensorM},
this equation is also written as
\begin{equation}
\ell_{a}=1-\frac{2}{N}\sum_{i}tr\left[P_{i}Q_{i}^{T}\right]+\frac{1}{N}\sum_{i}\sum_{l}\left(\left(tr\left[P_{l}Q_{i}^{T}\right]\right)^{2}+tr\left[P_{l}Q_{i}^{T}\left(L\cdot Q_{i}P_{l}^{T}+P_{l}Q_{i}^{T}\right)\right]\right).\label{eq_la_PQ_full}
\end{equation}
Here we defined $N\times N$ matrices $\left\{ P_{i}\right\} _{i=1}^{N}$
and $\left\{ Q_{i}\right\} _{i=1}^{N}$ by $\left[P_{i}\right]_{jk}=P_{ijk}$
and $\left[Q_{i}\right]_{jk}=Q_{ijk}$ as in the main text. By taking
the gradient with respect to $P_{l}$ and $Q_{i}$, we get 
\begin{equation}
\begin{aligned}\frac{\partial\ell_{a}}{\partial P_{l}}=0 & \Leftrightarrow Q_{l}=\sum_{i}\left(tr\left[P_{l}Q_{i}^{T}\right]I_{N}+L\cdot P_{l}Q_{i}^{T}+Q_{i}P_{l}^{T}\right)Q_{i},\\
\frac{\partial\ell_{a}}{\partial Q_{i}}=0 & \Leftrightarrow P_{i}=\sum_{l}\left(tr\left[P_{l}Q_{i}^{T}\right]I_{N}+L\cdot Q_{i}P_{l}^{T}+P_{l}Q_{i}^{T}\right)P_{l},
\end{aligned}
\label{eq_dla_dPl_dla_dQi}
\end{equation}
where $I_{N}$ is the size $N$ identity matrix. Under $P=Q$ constraint,
the above equations are rewritten as
\begin{equation}
P_{i}=\sum_{l}\left(tr\left[P_{l}P_{i}^{T}\right]I_{N}+L\cdot P_{i}P_{l}^{T}+P_{l}P_{i}^{T}\right)P_{l}.\label{eq_dla_dPi_sym}
\end{equation}

On the other hand, as a function of $\left\{ P_{i}\right\} $ and
$\left\{ R_{i}\right\} $, the decoding error for $\boldsymbol{b}$
is given as 
\begin{equation}
\ell_{b}=1-\frac{2}{N}\sum_{i}tr\left[P_{i}R_{i}^{T}\right]+\frac{1}{N}\sum_{i}\sum_{l}\left(tr\left[P_{i}R_{i}^{T}R_{l}P_{l}^{T}\right]+tr\left[P_{l}R_{i}^{T}\left(L\cdot R_{i}P_{l}^{T}+R_{l}P_{i}^{T}\right)\right]\right).\label{eq_lb_PR}
\end{equation}
Taking the gradient with respect to $P$ and $R$ , we get 
\begin{equation}
\begin{aligned}\frac{\partial\ell_{b}}{\partial P_{l}}=0 & \Leftrightarrow R_{l}=\sum_{i}\left(L\cdot P_{l}R_{i}^{T}R_{i}+P_{i}\left[R_{i}^{T}R_{l}+R_{l}^{T}R_{i}\right]\right),\\
\frac{\partial\ell_{b}}{\partial R_{i}}=0 & \Leftrightarrow P_{i}=\sum_{l}\left(L\cdot R_{i}P_{l}^{T}P_{l}+R_{l}\left[P_{l}^{T}P_{i}+P_{i}^{T}P_{l}\right]\right).
\end{aligned}
\label{eq_dlb_dPl_dlb_dRi}
\end{equation}
In particular, under $P=R$ constraint, the fixed-point condition
on $\left\{ P_{i}\right\} $ with respect to $\ell_{b}$ is given
as
\begin{equation}
P_{i}=\sum_{l}\left(L\cdot P_{i}P_{l}^{T}P_{l}+P_{l}P_{i}^{T}P_{l}+P_{l}P_{l}^{T}P_{i}\right),\label{eq_dlb_dPi_sym}
\end{equation}
for $i=1,...,N$. 

\subsection{Details of the numerical optimization algorithm \label{subsec:numerical_opt}}

Rewriting the fixed-point condition of $P_{l}$ with respect to the
loss $\ell_{a}$ (Eq. \ref{eq_dla_dPl_dla_dQi}), we get
\begin{equation}
Q_{ljk}=\sum_{m}\sum_{n}\sum_{i}\left(\delta_{jm}L\sum_{\beta}Q_{i\beta k}Q_{i\beta n}+Q_{ijk}Q_{imn}+Q_{ijn}Q_{imk}\right)P_{lmn},
\end{equation}
for $j,k=1,...,N$. Thus, introducing an $N^{2}\times N^{2}$ matrix
$\Gamma^{q}$ by
\begin{equation}
\Gamma_{[jN+k],[mN+n]}^{q}\equiv\sum_{i}\left(\delta_{jm}L\sum_{\beta}Q_{i\beta k}Q_{i\beta n}+Q_{ijk}Q_{imn}+Q_{ijn}Q_{imk}\right),
\end{equation}
we get $Q_{ljk}=\sum_{m,n}\Gamma_{[jN+k],[mN+n]}^{q}P_{lmn}$. Therefore,
we can derive the binding tensor $P$ that minimizes the loss $\ell_{a}$
under a fixed $Q$ by solving this linear equation as 
\begin{equation}
\text{Vec}\left[P_{l}\right]=\left(\Gamma^{q}\right)^{-1}\text{Vec}\left[Q_{l}\right],
\end{equation}
where $\text{Vec}\left[P_{l}\right]$ and $\text{Vec}\left[Q_{l}\right]$
are the vector representations of $N\times N$ matrices $P_{l}$ and
$Q_{l}$, respectively. From a similar calculation, the unbinding
tensor $Q$ that minimizes the loss $\ell_{a}$ under a fixed $P$
is given by
\begin{equation}
\text{Vec}\left[Q_{l}\right]=\left(\Gamma^{pa}\right)^{-1}\text{Vec}\left[P_{l}\right],
\end{equation}
where $\Gamma^{pa}$ is an $N^{2}\times N^{2}$ matrix:
\begin{equation}
\Gamma_{[jN+k],[mN+n]}^{pa}\equiv\sum_{i}\left(\delta_{jm}L\sum_{\beta}P_{i\beta k}P_{i\beta n}+P_{ijk}P_{imn}+P_{ijn}P_{imk}\right).
\end{equation}
To consider minimization of $\ell_{b}$, we rewrite the fixed-point
condition of $P_{i}$ with respect to $\ell_{b}$ (Eq. \ref{eq_dlb_dPl_dlb_dRi})
as
\begin{equation}
R_{lmk}=\sum_{i}\sum_{n}\Gamma_{lkin}^{r}P_{imn},
\end{equation}
where
\begin{equation}
\Gamma_{[lN+k],[iN+n]}^{r}\equiv\delta_{il}L\sum_{\alpha}\sum_{j}R_{\alpha jn}R_{\alpha jk}+\sum_{j}R_{ljk}R_{ijn}+\sum_{j}R_{ljn}R_{ijk}\label{eq_def_Gamma_r}
\end{equation}
is an $N^{2}\times N^{2}$ matrix. Let us construct $N\times N$ matrices
$\check{R}_{m}$ and $\check{P}_{m}$ by $\left[\check{R}_{m}\right]_{l,k}=R_{lmk}$
and $\left[\check{P}_{m}\right]_{l,k}=P_{lmk}$ respectively (for
$m=1,...,N$). Then, for a given $R$, the tensor $P$ that satisfies
the fixed-point condition of $\ell_{b}$ is derived as
\begin{equation}
\text{Vec}\left[\check{P}_{m}\right]=\left(\Gamma^{r}\right)^{-1}\text{Vec}\left[\check{R}_{m}\right]
\end{equation}
for $m=1,...,N$. Similarly, the fixed-point condition for $R$ is
written as
\begin{equation}
P_{ijk}=\sum_{l}\sum_{n}\Gamma_{ikln}^{pb}R_{ljn}
\end{equation}
where
\begin{equation}
\Gamma_{[iN+k],[lN+n]}^{pb}\equiv\delta_{il}L\sum_{\alpha}\sum_{m}P_{\alpha mk}P_{\alpha mn}+\sum_{m}P_{imk}P_{lmn}+\sum_{m}P_{imn}P_{lmk}.\label{eq_def_Gamma_pb}
\end{equation}
Therefore, for a given $P$, $R$ should satisfy
\begin{equation}
\text{Vec}\left[\check{R}_{j}\right]=\left(\Gamma^{pb}\right)^{-1}\text{Vec}\left[\check{P}_{j}\right]
\end{equation}
for $j=1,...,N$.

Combining the fixed-point algorithms for $P,Q$ in the main text,
and for $P,R$ described above, we obtain an iterative optimization
algorithm of $P,Q,R$. In a pseudocode, this algorithm is written
as Algorithm 1. In Figures 2 and 3, we initialized $P,Q,R$ randomly
by setting their elements from i.i.d Gaussian with variance $\frac{1}{NN_{c}}$,
then performed Algorithm 1 for $T=100$ iterations under $L=1$. In
Figure 5, we instead initialized $P,Q,R$ as a sparse $K$-compositional
binding tensor plus an element-wise Gaussian perturbation with variance
$\sigma^{2}/N^{2}$. Noise was added to $P,Q,R$ independently (hence
$P\neq Q\neq R$ after the perturbation). Figure 6
describes the optimization process under $L=3$. The optimization
process becomes slower and the advantage over HRR gets smaller (compare
Fig. 6A with Fig. 2A). However, the obtained binding
matrices exhibit $8\times8$ block diagonal structures when projected
to the space where $\bar{P}_{1}$ is diagonal (Figs. 6B
and C).

\subsection{Lower bound on the cardinality dependence\label{subsec:Lower-bound}}

Taking the expectation over $\boldsymbol{b}$, the readout (Eq. \ref{eq_def_qunbind_a})
becomes
\begin{equation}
\left\langle \hat{a}_{i}^{1}\right\rangle _{p(b)}=\left\langle \sum_{\mu}\sum_{j}\sum_{l}\sum_{m}\left(\sum_{k}Q_{ijk}P_{lmk}\right)b_{j}^{1}b_{m}^{\mu}a_{j}^{\mu}\right\rangle _{p(b)}=\sum_{j}\sum_{l}\left(\sum_{k}Q_{ijk}P_{ljk}\right)a_{l}^{1}.\label{eq_unbinding_amplitude}
\end{equation}
Thus, in order to retain the amplitude of the signal in the readout,
$P$ and $Q$ need to satisfy 
\begin{equation}
\sum_{j=1}^{N}\sum_{k=1}^{N_{c}}P_{ijk}Q_{ijk}=1.\label{eq_PQ_tr_constraint}
\end{equation}
Note that P and Q that minimize $\text{\ensuremath{\ell_{a}}}$ do
not necessarily satisfy this condition. However, the readout becomes
unbiased against the original signal under this condition (i.e., $\left\langle \hat{a}_{i}^{1}-a_{i}^{1}\right\rangle =0$).
Under this constraint, Eq. \ref{eq_la_PQ_full} is rewritten as
\begin{equation}
\ell_{a}=\frac{1}{N}\sum_{i}\sum_{l\neq i}\left(tr\left[P_{l}Q_{i}^{T}\right]\right)^{2}+\frac{1}{N}\sum_{i}\sum_{l}tr\left[P_{l}Q_{i}^{T}\left(P_{l}Q_{i}^{T}+LQ_{i}P_{l}^{T}\right)\right].
\end{equation}
Therefore at the large $L$ limit, the last term, $tr\left[P_{l}Q_{i}^{T}Q_{i}P_{l}^{T}\right]$,
becomes the dominant factor of the loss function. Hence, we consider
minimization of this dominant term under the constraint Eq. \ref{eq_PQ_tr_constraint}.
The Lagrangian for this constrained minimization is given by
\begin{equation}
\mathscr{\mathcal{L}}_{a}=\frac{1}{2}\sum_{i=1}^{N}\sum_{l=1}^{N}tr\left[P_{l}Q_{i}^{T}Q_{i}P_{l}^{T}\right]-\sum_{i=1}^{N}\lambda_{i}\left(tr\left[P_{i}Q_{i}^{T}\right]-1\right),
\end{equation}
where $\lambda_{i}\geq0$ is a Lagrange multiplier. The minimizer
$Q$ needs to satisfy
\begin{equation}
\frac{\partial\mathfrak{\mathcal{L}}_{a}}{\partial Q_{i}}=0\Leftrightarrow\lambda_{i}P_{i}=Q_{i}\sum_{l=1}^{N}P_{l}^{T}P_{l},\label{eq_dLa_Qi_LL}
\end{equation}
for $i=1,...,N$. Let us assume that $\sum_{l=1}^{N}P_{l}^{T}P_{l}$
is invertible. Note that, $\sum_{l=1}^{N}P_{l}^{T}P_{l}$ might not
be invertible especially under $N\ll N_{c}$, because $P_{l}$ is
a $N\times N_{c}$ matrix. However, because $\sum_{l=1}^{N}P_{l}^{T}P_{l}$
is a positive semi-definite matrix, if it is not invertible, the composition
$\boldsymbol{c}$ spans a subspace of $N_{c}$ dimensional space,
which should not provide any advantage over smaller $N_{c}$. Under
the invertibility assumption, $Q_{i}$ should satisfy 
\begin{equation}
Q_{i}=\lambda_{i}P_{i}\left(\sum_{l=1}^{N}P_{l}^{T}P_{l}\right)^{-1}.\label{eq_optQi_Lagrange}
\end{equation}
Substituting $Q_{i}$ in Eq. \ref{eq_PQ_tr_constraint} with the equation
above, we get
\begin{equation}
1=tr\left[P_{i}Q_{i}^{T}\right]=\lambda_{i}tr\left[P_{i}^{T}P_{i}\left({\textstyle \sum_{l=1}^{N}}P_{l}^{T}P_{l}\right)^{-1}\right].
\end{equation}
Because $P_{i}\left(\sum_{l}P_{l}^{T}P_{l}\right)^{-1}P_{i}^{T}$
is a positive semi-definite matrix, $tr\left[P_{i}^{T}P_{i}\left(\sum_{l}P_{l}^{T}P_{l}\right)^{-1}\right]>0$
(If all the eigenvalues are zero, the equation above does not hold).
Thus,
\begin{equation}
\lambda_{i}=\left(tr\left[P_{i}^{T}P_{i}\left({\textstyle \sum_{l=1}^{N}}P_{l}^{T}P_{l}\right)^{-1}\right]\right)^{-1}.
\end{equation}
Multiplying Eq. \ref{eq_dLa_Qi_LL} with $Q_{i}^{T}$ from the right,
taking the trace, and summing over $i$,
\begin{equation}
\sum_{i=1}^{N}\lambda_{i}tr\left[P_{i}Q_{i}^{T}\right]=\sum_{i=1}^{N}tr\left[Q_{i}^{T}Q_{i}\sum_{l=1}^{N}P_{l}^{T}P_{l}\right].
\end{equation}
Therefore,
\begin{equation}
\begin{aligned}\sum_{i=1}^{N}\sum_{l=1}^{N}tr\left[P_{l}Q_{i}^{T}Q_{i}P_{l}^{T}\right] & =\sum_{i=1}^{N}\lambda_{i}\\
 & =\sum_{i=1}^{N}\left(tr\left[P_{i}^{T}P_{i}\left({\textstyle \sum_{l=1}^{N}}P_{l}^{T}P_{l}\right)^{-1}\right]\right)^{-1}\\
 & \geq\frac{N^{2}}{\sum_{i=1}^{N}tr\left[P_{i}^{T}P_{i}\left({\textstyle \sum_{l=1}^{N}}P_{l}^{T}P_{l}\right)^{-1}\right]}=\frac{N^{2}}{N_{c}}.
\end{aligned}
\label{eq_Jensen_ineq}
\end{equation}
In the last line, we used Jensen's inequality with $1/x$ $\left(ie.\left(\frac{1}{N}\sum_{i}x_{i}\right)^{-1}\leq\frac{1}{N}\sum_{i}\frac{1}{x_{i}}\right)$.
Therefore, as long as $\sum_{l=1}^{N}P_{l}^{T}P_{l}$ is invertible,
the dominant error term of $\ell_{a}$ is lower bounded by $LN/N_{c}$.

\subsection{Decoding with a dictionary\label{subsec:dictionary_method}}

Although decoding in vector symbolic architecture is typically noisy
especially when $L>N_{c}/N$, if we know the dictionary of vectors
from which $\boldsymbol{a}$ is sampled, it is possible to recover
$\boldsymbol{a}$ accurately by matching the decoded vector $\hat{\boldsymbol{a}}$
with vectors $\left\{ \boldsymbol{a}_{\mu}\right\} _{\mu=1}^{D}$
in the dictionary. Below, we set $\mu=1$ as the target vector (as
before), $\mu=2,...,L$ as the other vectors bound to the composition
$\boldsymbol{c}$, and $\mu=L+1,...,D$ as the rest of vectors in
the dictionary. We denote the inner product between the retrieved
vector $\hat{\boldsymbol{a}}_{1}$ and $\mu$-th vector in the dictionary
$\boldsymbol{a}_{\mu}$ as
\begin{equation}
z_{\mu}\equiv\hat{\boldsymbol{a}}_{1}^{T}\boldsymbol{a}_{\mu},
\end{equation}
then pick $\boldsymbol{a}_{\mu}$ with the largest $z_{\mu}$ as the
decoded vector. Because all $\boldsymbol{a}_{\mu}$ has nearly the
same norm ($\left\Vert \boldsymbol{a}_{\mu}\right\Vert ^{2}\approx N$),
this is approximately equivalent to choosing $\boldsymbol{a}_{\mu}$
closest to $\hat{\boldsymbol{a}}_{1}$ in term of L2-norm (i.e., $\textrm{argmin}_{\mu}\left\Vert \hat{\boldsymbol{a}}_{1}-\boldsymbol{a}_{\mu}\right\Vert ^{2}$).
The probability of a correct classification under this decoding method
is
\begin{equation}
P_{\textrm{correct}}=\int dz_{1}...dz_{D}P\left[z_{1,}...,z_{D}\right]\hat{\theta}\left[z_{1}>z_{2},...,z_{1}>z_{D}\right],
\end{equation}
where $\hat{\theta}\left[x\right]$ is the indicator function. Because
$\left\{ z_{\mu}\right\} $ are not independent with each other, $P_{correct}$
is generally not analytically tractable. However, we can approximately
estimate $P_{\mathrm{correct}}$ under $N\gg1$ for any quadratic
binding methods satisfying Eq. \ref{eq_PQ_tr_constraint} in a similar
manner to previous works \citep{murdock1982theory,plate1995holographic,steinberg2022associative}.
We first normalize variables $\left\{ z_{\mu}\right\} _{\mu=1}^{D}$
as
\begin{equation}
\hat{z}_{\mu}\equiv\frac{\hat{\boldsymbol{a}}_{1}^{T}\boldsymbol{a}_{\mu}}{\boldsymbol{a}_{1}^{T}\boldsymbol{a}_{1}}.
\end{equation}
This normalization improves the accuracy of the Gaussian approximation
we introduce below. Because the normalization does not change the
relative order among $\left\{ z_{\mu}\right\} _{\mu=1}^{D}$, the
probability of a correct classification is written as
\begin{equation}
P_{\textrm{correct}}=\int d\hat{z}_{1}...d\hat{z}_{D}P\left[\hat{z}_{1},...,\hat{z}_{D}\right]\hat{\theta}\left[\hat{z}_{1}>\hat{z}_{2},...,\hat{z}_{1}>\hat{z}_{D}\right]
\end{equation}
To evaluate this integral, we approximate the probability distribution
$P\left[\hat{z}_{1},...,\hat{z}_{D}\right]$ with a Gaussian distribution:
\begin{equation}
P\left[\hat{z}_{1},...,\hat{z}_{D}\right]\approx q\left[\hat{z}_{1},...,\hat{z}_{D}\right]\equiv N\left(\boldsymbol{z};\bar{\boldsymbol{z}},\Sigma\right),
\end{equation}
where $\boldsymbol{z}=[\hat{z}_{1},...,\hat{z}_{D}]^{T}$. By definition,
$\hat{z}_{\mu}$ follows
\begin{equation}
\hat{z}_{\mu}=\frac{1}{\sum_{k=1}^{N}\left(a_{k}^{1}\right)^{2}}\sum_{\nu=1}^{L}\sum_{i=1}^{N}\sum_{j=1}^{N}\sum_{l=1}^{N}\sum_{m=1}^{N}M_{mj}^{li}a_{i}^{\mu}a_{l}^{\nu}b_{j}^{1}b_{m}^{\nu},
\end{equation}
where $M_{mj}^{li}\equiv\sum_{k}P_{lmk}Q_{ijk}$ (Eq. \ref{eq_def_tensorM}).
Taking the expectation over randomly sampled $\left\{ \boldsymbol{a}_{\mu}\right\} _{\mu=1}^{D}$
and $\left\{ \boldsymbol{b}_{\mu}\right\} _{\mu=1}^{L}$, the mean
is estimated as
\[
\left\langle \hat{z}_{\mu}\right\rangle =\left\langle \frac{\delta_{1\mu}}{\sum_{k=1}^{N}\left(a_{k}^{1}\right)^{2}}\sum_{i}\sum_{j}M_{jj}^{ii}\left(a_{i}^{1}\right)^{2}\right\rangle =\delta_{1\mu}
\]
 Here, we used Eq. \ref{eq_PQ_tr_constraint}. On the other hand,
the covariance becomes
\begin{equation}
\begin{aligned}\Sigma_{\mu\nu} & =\left\langle \left(\hat{z}_{\mu}-\delta_{1\mu}\right)\left(\hat{z}_{\nu}-\delta_{1\nu}\right)\right\rangle \\
 & =\sum_{\rho,\sigma}\sum_{i,i^{\prime}}\sum_{j,j^{\prime}}\sum_{l,l^{\prime}}\sum_{m,m^{\prime}}M_{mj}^{li}M_{m^{\prime}j^{\prime}}^{l^{\prime}i^{\prime}}\left\langle b_{j}^{1}b_{m}^{\rho}b_{j^{\prime}}^{1}b_{m^{\prime}}^{\sigma}\right\rangle \left\langle \frac{1}{\left(\sum_{k}a_{k}^{1}a_{k}^{1}\right)^{2}}a_{l}^{\rho}a_{i}^{\mu}a_{l^{\prime}}^{\sigma}a_{i^{\prime}}^{\nu}\right\rangle -\delta_{1\mu}\delta_{1\nu},
\end{aligned}
\label{eq_zhat_covariance}
\end{equation}
The expectation over $\boldsymbol{b}$ is non-zero only when $\rho=\sigma$,
but given $\rho=\sigma$, the expectation over $\boldsymbol{a}$ is
non-zero only when $\mu=\nu$. Therefore, $\Sigma_{\mu\nu}=0$ for
$\mu\neq\nu$, meaning that the joint distribution $q\left[\hat{z}_{1},...,\hat{z}_{D}\right]$
is factorized under the Gaussian approximation. The second moment
is evaluated as
\begin{equation}
\left\langle \left(\hat{z}_{\mu}\right)^{2}\right\rangle =\sum_{\rho=1}^{L}\sum_{i,i^{\prime}}\sum_{j,j^{\prime}}\sum_{l,l^{\prime}}\sum_{m,m^{\prime}}M_{mj}^{li}M_{m^{\prime}j^{\prime}}^{l^{\prime}i^{\prime}}\left\langle b_{j}^{1}b_{j^{\prime}}^{1}b_{m}^{\rho}b_{m^{\prime}}^{\rho}\right\rangle \left\langle \frac{1}{\left(\sum_{k}a_{k}^{1}a_{k}^{1}\right)^{2}}a_{i}^{\mu}a_{i^{\prime}}^{\mu}a_{l}^{\rho}a_{l^{\prime}}^{\rho}\right\rangle .
\end{equation}
The expectation over $\boldsymbol{b}$ is given as
\begin{equation}
\left\langle b_{j}^{1}b_{j^{\prime}}^{1}b_{m}^{\rho}b_{m^{\prime}}^{\rho}\right\rangle =\delta_{jj^{\prime}}\delta_{mm^{\prime}}+\delta_{1\rho}\left[\delta_{jm}\delta_{j^{\prime}m^{\prime}}+\delta_{jm^{\prime}}\delta_{j^{\prime}m}\right],
\end{equation}
while the expectation over $\boldsymbol{a}$ is estimated as (see
Appendix \ref{subsec_expectation_a})
\begin{equation}
\left\langle \frac{1}{\left(\sum_{k}a_{k}^{1}a_{k}^{1}\right)^{2}}a_{i}^{\mu}a_{i^{\prime}}^{\mu}a_{l}^{\rho}a_{l^{\prime}}^{\rho}\right\rangle =\begin{cases}
\frac{\delta_{ii'}\delta_{ll'}+\delta_{\mu\rho}\left(\delta_{il}\delta_{i'l'}+\delta_{il'}\delta_{i'l}\right)}{\left(N-2\right)\left(N-4\right)} & \mu\geq2\\
\frac{\delta_{ii'}\delta_{ll'}}{N\left(N-2\right)}+\frac{\delta_{1\rho}\left(\delta_{il}\delta_{i'l'}+\delta_{il'}\delta_{i'l}\right)}{N\left(N+2\right)} & \mu=1.
\end{cases}
\end{equation}
Thus, for $\mu>2$, the variance $\left\langle \left(\hat{z}_{\mu}\right)^{2}\right\rangle $
follows
\begin{equation}
\left\langle \left(z_{\mu}\right)^{2}\right\rangle =\sigma_{o}^{2}+\sigma_{1}^{2}+\left[\mu\leq L\right]_{+}\sigma_{2}^{2},
\end{equation}
where 
\begin{equation}
\begin{aligned}\sigma_{0}^{2} & \equiv\frac{1}{\left(N-2\right)\left(N-4\right)}\sum_{\rho}\sum_{i,i^{\prime}}\sum_{j,j^{\prime}}\sum_{l,l^{\prime}}\sum_{m,m^{\prime}}M_{mj}^{li}M_{m^{\prime}j^{\prime}}^{l^{\prime}i^{\prime}}\delta_{jj^{\prime}}\delta_{mm^{\prime}}\delta_{ii^{\prime}}\delta_{ll^{\prime}},\\
\sigma_{1}^{2} & \equiv\frac{1}{\left(N-2\right)\left(N-4\right)}\sum_{i,i^{\prime}}\sum_{j,j^{\prime}}\sum_{l,l^{\prime}}\sum_{m,m^{\prime}}M_{mj}^{li}M_{m^{\prime}j^{\prime}}^{l^{\prime}i^{\prime}}\left[\delta_{jm}\delta_{j'm'}+\delta_{jm'}\delta_{j'm}\right]\delta_{ii'}\delta_{ll'},\\
\sigma_{2}^{2} & \equiv\frac{1}{\left(N-2\right)\left(N-4\right)}\sum_{i,i^{\prime}}\sum_{j,j^{\prime}}\sum_{l,l^{\prime}}\sum_{m,m^{\prime}}M_{mj}^{li}M_{m^{\prime}j^{\prime}}^{l^{\prime}i^{\prime}}\delta_{jj^{\prime}}\delta_{mm^{\prime}}\left[\delta_{il}\delta_{i^{\prime}l^{\prime}}+\delta_{il^{\prime}}\delta_{i^{\prime}l}\right],
\end{aligned}
\end{equation}
Note that, $\sigma_{2}^{2}$ term appears only when the vector $\boldsymbol{a}_{\mu}$
is bound to the composition $\boldsymbol{c}$ (ie, $\mu\leq L$).
Summing over the delta functions and using $N\gg1$, the components
$\sigma_{0}^{2}$, $\sigma_{1}^{2}$, $\sigma_{2}^{2}$ are rewritten
as
\begin{equation}
\begin{aligned}\sigma_{0}^{2} & =\frac{L}{N^{2}}\sum_{i,l}\sum_{j,m}\left(M_{mj}^{li}\right)^{2},\\
\sigma_{1}^{2} & =\frac{1}{N^{2}}\sum_{i,l}\left(\left[\sum_{j}M_{jj}^{li}\right]^{2}+\sum_{j,m}M_{mj}^{li}M_{jm}^{li}\right),\\
\sigma_{2}^{2} & =\frac{1}{N^{2}}\sum_{j,m}\left(\left[\sum_{i}M_{mj}^{ii}\right]^{2}+\sum_{i,l}M_{mj}^{li}M_{mj}^{il}\right),
\end{aligned}
\label{eq_def_sigma0_2}
\end{equation}
On the other hand, at $\mu=1$, the variance follows 
\begin{equation}
\left\langle \left(z_{\mu}-\delta_{1\mu}\right)^{2}\right\rangle \approx\sigma_{o}^{2}+\sigma_{1}^{2}+\sigma_{2}^{2}+\sigma_{3}^{2},
\end{equation}
where the extra term $\sigma_{3}^{2}$ is given as
\begin{equation}
\begin{aligned}\sigma_{3}^{2} & \equiv\frac{1}{N\left(N+2\right)}\sum_{i,i^{\prime}}\sum_{j,j^{\prime}}\sum_{l,l^{\prime}}\sum_{m,m^{\prime}}M_{mj}^{li}M_{m^{\prime}j^{\prime}}^{l^{\prime}i^{\prime}}\left(\delta_{il}\delta_{i^{\prime}l^{\prime}}+\delta_{il^{\prime}}\delta_{i^{\prime}l}\right)\left(\delta_{jm}\delta_{j'm'}+\delta_{jm'}\delta_{j'm}\right)-1\\
 & =\frac{1}{N\left(N+2\right)}\left(\sum_{i,j}M_{jj}^{ii}\right)^{2}-1+\frac{1}{N\left(N+2\right)}\sum_{i,i^{\prime}}\sum_{j,j^{\prime}}\sum_{l,l^{\prime}}\sum_{m,m^{\prime}}M_{mj}^{li}M_{m^{\prime}j^{\prime}}^{l^{\prime}i^{\prime}}\left(\delta_{il}\delta_{i^{\prime}l^{\prime}}\delta_{jm'}\delta_{j'm}+\delta_{il^{\prime}}\delta_{i^{\prime}l}\left[\delta_{jm}\delta_{j'm'}+\delta_{jm'}\delta_{j'm}\right]\right)\\
 & \approx\frac{-2}{N}+\frac{1}{N^{2}}\left(\sum_{j,m}\left[\sum_{i}M_{mj}^{ii}\right]\left[\sum_{l}M_{jm}^{ll}\right]+\sum_{i,l}\left[\sum_{j}M_{jj}^{li}\right]\left[\sum_{m}M_{mm}^{il}\right]+\sum_{i,l}\sum_{j,m}M_{mj}^{li}M_{jm}^{il}\right).
\end{aligned}
\end{equation}
Notably, of the four terms $\sigma_{0}^{2}$,...,$\sigma_{3}^{2}$
consist of the variance, only $\sigma_{0}^{2}$ scales with the number
of bound pairs $L$. Thus, at the large $L$ limit, the variance of
all $\mu$ follows 
\begin{equation}
\left\langle \left(\hat{z}_{\mu}-\delta_{1\mu}\right)^{2}\right\rangle =L\left(\sum_{i,l}\sum_{j,m}\left(M_{mj}^{li}\right)^{2}+\mathcal{O}\left(\frac{1}{L}\right)\right)\geq L\left(\frac{N^{2}}{N_{c}}+\mathcal{O}\left(\frac{1}{L}\right)\right).
\end{equation}
The last inequality follows from Eq. \ref{eq_Jensen_ineq}. On the
other hand, under $L\sim\mathcal{O}\left(1\right)$, $\sigma_{1}^{2}$,
$\sigma_{2}^{2}$, and $\sigma_{3}^{2}$ may play an important role.
For convenience, let us denote 
\begin{equation}
\sigma_{s}^{2}\equiv\sigma_{0}^{2}+\sigma_{1}^{2}+\sigma_{2}^{2}+\sigma_{3}^{2},\;\sigma_{l}^{2}\equiv\sigma_{0}^{2}+\sigma_{1}^{2}+\sigma_{2}^{2},\;\sigma_{d}^{2}\equiv\sigma_{0}^{2}+\sigma_{1}^{2}.
\end{equation}
The first term $\sigma_{s}^{2}$ corresponds to the variance of the
target readout $\hat{z}_{1}$, while $\sigma_{l}^{2}$ is the variance
of readout $\hat{z}_{\mu}$ for $\mu=2,...,L$, and $\sigma_{d}^{2}$
is the variance of $\hat{z}_{\mu}$ for $\mu=L+1,...,D$. Because
$q[\hat{z}_{1},...,\hat{z}_{D}]$ is factorized, using $\sigma_{s}^{2}$,
$\sigma_{l}^{2}$, $\sigma_{d}^{2}$, we get
\[
\begin{aligned}P_{\mathrm{correct}} & \approx\int d\hat{z}_{1}q\left[\hat{z}_{1}\right]\prod_{\mu=2}^{D}\int d\hat{z}_{\mu}q\left[\hat{z}_{\mu}\right]\theta\left[\hat{z}_{1}>\hat{z}_{\mu}\right]\\
 & =\int\frac{d\hat{z}_{1}}{\sqrt{2\pi\sigma_{s}^{2}}}e^{-\left(\hat{z}_{1}-1\right)^{2}/2\sigma_{s}^{2}}\left(\int_{-\infty}^{\hat{z}_{1}}\frac{dz_{l}}{\sqrt{2\pi\sigma_{l}^{2}}}e^{-z_{l}^{2}/2\sigma_{l}^{2}}\right)^{L-1}\left(\int_{-\infty}^{\hat{z}_{1}}\frac{dz_{d}}{\sqrt{2\pi\sigma_{d}^{2}}}e^{-z_{d}^{2}/2\sigma_{d}^{2}}\right)^{D-L}.\\
 & =\int\frac{dz}{\sqrt{2\pi}}\exp\left[-\frac{1}{2}\left(z-\frac{1}{\sigma_{s}}\right)^{2}\right]\left(\Phi\left[\frac{\sigma_{s}z}{\sigma_{l}}\right]\right)^{L-1}\left(\Phi\left[\frac{\sigma_{s}z}{\sigma_{d}}\right]\right)^{D-L},
\end{aligned}
\]
where $\Phi[z]\equiv\int_{-\infty}^{z}\frac{dy}{\sqrt{2\pi}}e^{-y^{2}/2}$
is the cumulative distribution function of a Gaussian distribution
$\mathcal{N}(0,1)$. Here, $\left(\Phi\left[\frac{\sigma_{s}z}{\sigma_{l}}\right]\right)^{L-1}$
evaluates the probability of correct classification against vectors
in the composition $\boldsymbol{c}$, while $\left(\Phi\left[\frac{\sigma_{s}z}{\sigma_{d}}\right]\right)^{D-L}$
captures the classification accuracy against the rest of words in
the dictionary. 

Because $\sigma_{s}^{2},\sigma_{l}^{2},\sigma_{d}^{2}$ depend on
the choice of the binding and unbinding operators $P$ and $Q$, $P_{\textrm{correct}}$
also depends on the binding methods. For instance, under the octonion
binding, from Eq. \ref{eq_def_sigma0_2} and Eq. \ref{eq_def_sparse_Kcomp},
we get 
\begin{equation}
\sigma_{s}^{2}=\frac{1}{N}\left(L+\frac{3}{2}\right),\quad\sigma_{l}^{2}=\frac{1}{N}\left(L+\frac{1}{2}\right),\quad\sigma_{d}^{2}=\frac{1}{N}\left(L+\frac{1}{4}\right),
\end{equation}
whereas under HRR and also under the random binding, assuming $N\gg1$,
\begin{equation}
\sigma_{s}^{2}=\frac{1}{N}\left(L+3\right),\quad\sigma_{l}^{2}=\frac{1}{N}\left(L+2\right),\quad\sigma_{d}^{2}=\frac{1}{N}\left(L+1\right).
\end{equation}

\subsection{Estimation of $\boldsymbol{a}$-dependent terms in the variance\label{subsec_expectation_a}}

Here, we estimate the expectation of $\nicefrac{a_{i}^{\mu}a_{i^{\prime}}^{\mu}a_{l}^{\rho}a_{l^{\prime}}^{\rho}}{\left(\sum_{k=1}^{N}a_{k}^{1}a_{k}^{1}\right)^{2}}$
over random Gaussian vectors $\boldsymbol{a}$ under $N\gg1$. First,
for $\mu\geq2$, we get, 
\begin{equation}
\begin{aligned} & \left\langle \frac{1}{\left(\sum_{k=1}^{N}a_{k}^{1}a_{k}^{1}\right)^{2}}a_{i}^{\mu}a_{i^{\prime}}^{\mu}a_{l}^{\rho}a_{l^{\prime}}^{\rho}\right\rangle \\
 & =\left(\delta_{ii'}\delta_{ll'}+\delta_{\mu\rho}\left[\delta_{il}\delta_{i'l'}+\delta_{il'}\delta_{i'l}\right]\right)\left\langle \frac{1}{\left(\sum_{k}a_{k}^{1}a_{k}^{1}\right)^{2}}\right\rangle +\delta_{1\rho}\delta_{ii'}\delta_{ll'}\left(\left\langle \frac{a_{l}^{1}a_{l}^{1}}{\left(\sum_{k}a_{k}^{1}a_{k}^{1}\right)^{2}}\right\rangle -\left\langle \frac{1}{\left(\sum_{k}a_{k}^{1}a_{k}^{1}\right)^{2}}\right\rangle \right)\\
 & =\frac{1}{\left(N-2\right)\left(N-4\right)}\left(\delta_{ii'}\delta_{ll'}+\delta_{\mu\rho}\left[\delta_{il}\delta_{i'l'}+\delta_{il'}\delta_{i'l}\right]\right)-\frac{4\delta_{1\rho}\delta_{ii'}\delta_{ll'}}{N\left(N-2\right)\left(N-4\right)}.
\end{aligned}
\end{equation}
In the last line, we used 
\begin{equation}
\begin{aligned}\left\langle \frac{1}{\left(\sum_{k}a_{k}^{1}a_{k}^{1}\right)^{2}}\right\rangle _{a} & =\left\langle \frac{1}{x^{2}}\right\rangle _{x\sim\chi_{N}^{2}}=\frac{1}{\left(N-2\right)\left(N-4\right)},\\
\left\langle \frac{a_{l}^{1}a_{l}^{1}}{\left(\sum_{k}a_{k}^{1}a_{k}^{1}\right)^{2}}\right\rangle _{a} & =\left\langle \frac{x}{\left(x+y\right)^{2}}\right\rangle _{x\sim\chi_{1}^{2},y\sim\chi_{N-1}^{2}}=\frac{1}{N\left(N-2\right)},
\end{aligned}
\end{equation}
where $\chi_{k}^{2}$ is the chi-squared distribution with degree
$k$. Similarly, under $\mu=1$, 
\begin{equation}
\begin{aligned}\left\langle \frac{1}{\left(\sum_{k}a_{k}^{1}a_{k}^{1}\right)^{2}}a_{i}^{\mu}a_{i^{\prime}}^{\mu}a_{l}^{\rho}a_{l^{\prime}}^{\rho}\right\rangle  & \approx\left(1-\delta_{1\rho}\right)\delta_{ii'}\delta_{ll'}\left\langle \frac{a_{l}^{1}a_{l}^{1}}{\left(\sum_{k}a_{k}^{1}a_{k}^{1}\right)^{2}}\right\rangle +\delta_{1\rho}\left(\delta_{ii'}\delta_{ll'}+\delta_{il}\delta_{i'l'}+\delta_{il'}\delta_{i'l}\right)\left\langle \frac{a_{i}^{1}a_{i}^{1}a_{l}^{1}a_{l}^{1}}{\left(\sum_{k}a_{k}^{1}a_{k}^{1}\right)^{2}}\right\rangle \\
 & =\frac{\left(1-\delta_{1\rho}\right)\delta_{ii'}\delta_{ll'}}{N\left(N-2\right)}+\frac{\delta_{1\rho}\left(\delta_{ii'}\delta_{ll'}+\delta_{il}\delta_{i'l'}+\delta_{il'}\delta_{i'l}\right)}{N\left(N+2\right)}.
\end{aligned}
\end{equation}
In the last line, we used
\begin{equation}
\left\langle \frac{a_{i}^{1}a_{i}^{1}a_{l}^{1}a_{l}^{1}}{\left(\sum_{k}a_{k}^{1}a_{k}^{1}\right)^{2}}\right\rangle =\left\langle \frac{xy}{\left(x+y+z\right)^{2}}\right\rangle _{x,y\sim\chi_{1}^{2},z\sim\chi_{N-2}^{2}}=\frac{1}{N\left(N+2\right)}.
\end{equation}
Therefore, up to the leading order terms, 
\begin{equation}
\left\langle \frac{1}{\left(\sum_{k}a_{k}^{1}a_{k}^{1}\right)^{2}}a_{i}^{\mu}a_{i^{\prime}}^{\mu}a_{l}^{\rho}a_{l^{\prime}}^{\rho}\right\rangle =\begin{cases}
\frac{\delta_{ii'}\delta_{ll'}+\delta_{\mu\rho}\left(\delta_{il}\delta_{i'l'}+\delta_{il'}\delta_{i'l}\right)}{\left(N-2\right)\left(N-4\right)} & \mu\geq2\\
\frac{\delta_{ii'}\delta_{ll'}}{N\left(N-2\right)}+\frac{\delta_{1\rho}\left(\delta_{il}\delta_{i'l'}+\delta_{il'}\delta_{i'l}\right)}{N\left(N+2\right)} & \mu=1.
\end{cases}
\end{equation}

\subsection{Performance of the random binding\label{subsec:random_binding}}

We construct a random binding tensor by setting $P=Q=R$, and choosing
their elements from i.i.d Gaussian with the mean zero and the variance
$\frac{1}{NN_{c}}$. Under this normalization, $P$ satisfies Eq.
\ref{eq_PQ_tr_constraint} under $N,N_{c}\gg1$. From Eq. \ref{eq_la_full},
the average error over randomly chosen $P$ is given as
\begin{equation}
\ell_{a}=\frac{LN}{N_{c}}\left(1+\frac{N_{c}+1}{N^{2}}\right)+\left(\frac{1}{N}+\frac{3}{N_{c}}+\frac{1}{NN_{c}}\right).\label{eq_la_random}
\end{equation}
If $N_{c}\ll N^{2}$, the leading order term is $\ell_{a}\approx\frac{LN}{N_{c}}$,
which is the same with the lower bound. On the other hand, at $N_{c}=N^{2}$
limit, the leading order term becomes $\ell_{a}\approx\frac{2LN}{N_{c}}$,
which is twice larger than that of the tensor product representation
(Eq. \ref{eq_la_tenHRR}). 

\section{$K$-compositional binding and its extensions}

\subsection{Sufficiency of the Hurwitz matrix equations for the fixed-point condition\label{subsec:sufficiency}}

Here we show that Hurwitz matrix equations $P_{j}P_{i}^{T}+P_{i}P_{j}^{T}=2\lambda\delta_{ij}I_{N}$
with $\lambda=\frac{1}{LN+2}$ are sufficient for the fixed-point
conditions with respect to both $\ell_{a}$ (Eq. \ref{eq_dla_dPi_sym})
and $\ell_{b}$ (Eq. \ref{eq_dlb_dPi_sym}) under $P=Q=R$ constraint.
Recall that, $L$ is the number of bound pairs in the composition.
Although we mainly focused on $L=1$ in the main text, below we prove
the results for arbitrary $L$. Firstly, taking the trace of the Hurwitz
matrix equations, we get $tr\left[P_{i}P_{j}^{T}\right]=\lambda N\delta_{ij}$.
Moreover, because $P_{i}$ is a square matrix, $P_{i}P_{i}^{T}=\lambda I_{N}$
implies $P_{i}^{T}P_{i}=\lambda I_{N}$. Thus, 
\begin{equation}
\begin{aligned}\sum_{j=1}^{N}\left(tr\left[P_{j}P_{i}^{T}\right]I_{N}+LP_{i}P_{j}^{T}+P_{j}P_{i}^{T}\right)P_{j} & =\sum_{j=1}^{N}\left(\lambda N\delta_{ij}I_{N}+2\lambda\delta_{ij}I_{N}\right)P_{j}+\left(L-1\right)P_{i}\sum_{j=1}^{N}P_{j}^{T}P_{j}\\
 & =\lambda\left(LN+2\right)P_{i}=P_{i}.
\end{aligned}
\end{equation}
Secondly, using $\sum_{j}\left(P_{j}P_{i}^{T}+P_{i}P_{j}^{T}\right)P_{j}=2\lambda P_{i}$,
\begin{equation}
\begin{aligned}\sum_{j}\left(LP_{i}P_{j}^{T}P_{j}+P_{j}P_{i}^{T}P_{j}+P_{j}P_{j}^{T}P_{i}\right) & =\left(L-1\right)P_{i}\sum_{j=1}^{N}P_{j}^{T}P_{j}+\sum_{j=1}^{N}\left[P_{i}P_{j}^{T}+P_{j}P_{i}^{T}\right]P_{j}+\sum_{j=1}^{N}P_{j}P_{j}^{T}P_{i}\\
 & =\lambda\left(LN+2\right)P_{i}=P_{i}.
\end{aligned}
\end{equation}
Hence, a family of matrices $\left\{ P_{i}\right\} _{i=1}^{N}$ satisfying
Eq. \ref{eq_hurwitz_mat_eq} also satisfies Eqs. \ref{eq_dla_dPi_sym}
and \ref{eq_dlb_dPi_sym}. In particular, under $L=1$, it satisfies
Eqs. \ref{eq_dla_dp_sym} and \ref{eq_dlb_dp_sym}.

\subsection{Octonion binding\label{subsec:Octonion-binding}}

Using the Cayley-Dickson construction, a matrix representation of
an element $a=\left(a_{1},a_{2},...,a_{8}\right)$ of the octonion
algebra is given as \citep{tian2000matrix}
\begin{equation}
\phi(a)=\left(\begin{array}{cccccccc}
a_{1} & -a_{2} & -a_{3} & -a_{4} & -a_{5} & -a_{6} & -a_{7} & -a_{8}\\
a_{2} & a_{1} & a_{4} & -a_{3} & a_{6} & -a_{5} & -a_{8} & a_{7}\\
a_{3} & -a_{4} & a_{1} & a_{2} & a_{7} & a_{8} & -a_{5} & -a_{6}\\
a_{4} & a_{3} & -a_{2} & a_{1} & a_{8} & -a_{7} & a_{6} & -a_{5}\\
a_{5} & -a_{6} & -a_{7} & -a_{8} & a_{1} & a_{2} & a_{3} & a_{4}\\
a_{6} & a_{5} & -a_{8} & a_{7} & -a_{2} & a_{1} & -a_{4} & a_{3}\\
a_{7} & a_{8} & a_{5} & -a_{6} & -a_{3} & a_{4} & a_{1} & -a_{2}\\
a_{8} & -a_{7} & a_{6} & a_{5} & -a_{4} & -a_{3} & a_{2} & a_{1}
\end{array}\right).\;
\end{equation}
Because octonions are not associative under multiplication (i.e. there
are octonions $a,b,c,$ such that $a\cdot\left(b\cdot c\right)\neq\left(a\cdot b\right)\cdot c$),
a matrix representation of an octonion is not faithful, unlike matrix
representations of the quaternions and the complex numbers. However,
from $\phi(a)$, we can still construct a family of matrices $P=[P_{1},...,P_{8}]$
that satisfies the Hurwitz matrix equations which we can use as the
basis of binding matrices. Under this binding, up to a constant factor,
the composition $\boldsymbol{c}$ of two elements is calculated as
\[
\begin{aligned}c_{1}= & a_{1}b_{1}+a_{2}b_{2}+a_{3}b_{3}+a_{4}b_{4}+a_{5}b_{5}+a_{6}b_{6}+a_{7}b_{7}+a_{8}b_{8},\\
c_{2}= & a_{1}b_{2}-a_{2}b_{1}+a_{3}b_{4}-a_{4}b_{3}+a_{5}b_{6}-a_{6}b_{5}-a_{7}b_{8}+a_{8}b_{7},\\
c_{3}= & a_{1}b_{3}-a_{2}b_{4}-a_{3}b_{1}+a_{4}b_{2}+a_{5}b_{7}+a_{6}b_{8}-a_{7}b_{5}-a_{8}b_{6},\\
c_{4}= & a_{1}b_{4}+a_{2}b_{3}-a_{3}b_{2}-a_{4}b_{1}+a_{5}b_{8}-a_{6}b_{7}+a_{7}b_{6}-a_{8}b_{5},\\
c_{5}= & a_{1}b_{5}-a_{2}b_{6}-a_{3}b_{7}-a_{4}b_{8}-a_{5}b_{1}+a_{6}b_{2}+a_{7}b_{3}+a_{8}b_{4},\\
c_{6}= & a_{1}b_{6}+a_{2}b_{5}-a_{3}b_{8}+a_{4}b_{7}-a_{5}b_{2}-a_{6}b_{1}-a_{7}b_{4}+a_{8}b_{3},\\
c_{7}= & a_{1}b_{7}+a_{2}b_{8}+a_{3}b_{5}-a_{4}b_{6}-a_{5}b_{3}+a_{6}b_{4}-a_{7}b_{1}-a_{8}b_{2},\\
c_{8}= & a_{1}b_{8}-a_{2}b_{7}+a_{3}b_{6}+a_{4}b_{5}-a_{5}b_{4}-a_{6}b_{3}+a_{7}b_{2}-a_{8}b_{1}.
\end{aligned}
\]
Note that, because matrix representation of octonions is not unique,
there are various different ways to construct binding matrices that
have octonion structure. 

\subsection{Properties of the sparse $K$-compositional bindings \label{subsec:sparse_Kcomp}}

In Eq. \ref{eq_def_sparse_Kcomp}, we generated sparse $K$-compositional
binding operators by a block-wise binding. However, there are many
equivalent binding operators due to invariance. In particular, we
can generate a family of binding operators using an $N\times N$ orthogonal
matrix $W$ ($WW^{T}=W^{T}W=I_{N}$). Let us denoting $A=\{A_{1},...,A_{K}\}$
as a family of $K\times K$ matrices that satisfies the Hurwitz matrix
equations
\begin{equation}
A_{i}A_{j}^{T}+A_{j}A_{i}^{T}=2\lambda\delta_{ij}I_{K}\label{eq_Hurwitz_mat_norm}
\end{equation}
for $i,j=1,...,K$, with the normalization factor:
\begin{equation}
\lambda=\frac{1}{LK+2}.
\end{equation}
Setting $N=qK$ with a natural number $q$, we construct a binding
operator $P_{n}$ ($n=1,...,N$) by
\begin{equation}
P_{n}=\left(\sum_{k=1}^{K}W_{n,k}A_{k}\right)\oplus\left(\sum_{k=1}^{K}W_{n,K+k}A_{k}\right)\oplus...\oplus\left(\sum_{k=1}^{K}W_{n,(q-1)K+k}A_{k}\right).\label{eq_def_ortho_Kcomp_P}
\end{equation}
 In other words, we set the $r$-th block diagonal component of $P_{n}$
to $\sum_{k=1}^{K}W_{n,(r-1)K+k}A_{k}$. If we choose $W=I_{N}$,
we recover Eq. \ref{eq_def_sparse_Kcomp}. Below we show that under
this binding, for arbitrary positive integer $L,$ the decoding error
becomes $\ell_{a}=\ell_{b}=\frac{(L-1)K+2}{LK+2}$ and $\left\{ P_{n}\right\} _{n=1}^{N}$
satisfies the fixed-point conditions for both $\ell_{a}$ and $\ell_{b}$.
In particular, we recover $\ell_{a}=\ell_{b}=\frac{2}{K+2}$ under
$L=1$. 

\paragraph*{Decoding error of $\boldsymbol{a}$}

Here we show that, under this binding, the error $\ell_{a}$ (Eq.
\ref{eq_la_PQ_full}) satisfies $\ell_{a}=\frac{(L-1)K+2}{LK+2}$.
Firstly, using $tr[A_{k}A_{n}^{T}]=\lambda K\delta_{kn}$, $tr\left[P_{i}P_{l}^{T}\right]$
becomes
\begin{equation}
\begin{aligned}tr\left[P_{i}P_{l}^{T}\right] & =\sum_{r=0}^{q-1}tr\left[\left(\sum_{k=1}^{K}W_{i,rK+k}A_{k}\right)\left(\sum_{n=1}^{K}W_{l,rK+n}A_{n}^{T}\right)\right]\\
 & =\lambda K\sum_{r=0}^{q-1}\sum_{k=1}^{K}W_{i,rK+k}W_{l,rK+n}\\
 & =\lambda K\left[WW^{T}\right]_{il}=\lambda K\delta_{il}.
\end{aligned}
\label{eq_trPP_ortho}
\end{equation}
In the last line, we used the fact that $W$ is an orthogonal matrix.
Similarly, using $A_{i}A_{j}^{T}+A_{j}A_{i}^{T}=2\lambda\delta_{ij}I_{K}$,
\begin{equation}
\begin{aligned} & \sum_{i=1}^{N}\sum_{l=1}^{N}tr\left[P_{l}P_{i}^{T}\left(P_{l}P_{i}^{T}+P_{i}P_{l}^{T}\right)\right]\\
 & =\sum_{i=1}^{N}\sum_{l=1}^{N}\sum_{r=0}^{q-1}tr\left[\sum_{k=1}^{K}\sum_{k^{\prime}=1}^{K}W_{l,rK+k}W_{i,rK+k^{\prime}}A_{k}A_{k^{\prime}}^{T}\sum_{n=1}^{K}\sum_{n^{\prime}=1}^{K}W_{l,rK+n}W_{i,rK+n^{\prime}}\left(A_{n}A_{n^{\prime}}^{T}+A_{n^{\prime}}A_{n}^{T}\right)\right]\\
 & =2\lambda^{2}K\sum_{r=0}^{q-1}\sum_{k=1}^{K}\sum_{n=1}^{K}\left(\left[W^{T}W\right]_{rK+k,rK+n}\right)^{2}\\
 & =2\lambda^{2}NK.
\end{aligned}
\end{equation}
Finally, 
\begin{equation}
\begin{aligned}\sum_{i=1}^{N}\sum_{l=1}^{N}tr\left[P_{l}P_{i}^{T}P_{i}P_{l}^{T}\right] & =\sum_{r=0}^{q-1}\sum_{k=1}^{K}\sum_{m=1}^{K}\sum_{k'=1}^{K}\sum_{m'=1}^{K}\left[W^{T}W\right]_{rK+k,rK+k'}\left[W^{T}W\right]_{rK+m,rK+m'}tr\left[A_{k}A_{m}^{T}A_{m'}A_{k'}^{T}\right]\\
 & =\sum_{r=0}^{q-1}\sum_{k=1}^{K}\sum_{m=1}^{K}tr\left[A_{k}A_{m}^{T}A_{m}A_{k}^{T}\right]\\
 & =N\lambda^{2}K^{2}.
\end{aligned}
\end{equation}
Therefore, from Eq. \ref{eq_la_PQ_full}, the loss $\ell_{a}$ becomes
\begin{equation}
\ell_{a}=1-2\lambda K+\left(\lambda K\right)^{2}+2\lambda^{2}K+\left(L-1\right)\lambda^{2}K^{2}=\frac{(L-1)K+2}{LK+2}.
\end{equation}

Moreover, the binding operator defined by Eq. \ref{eq_def_ortho_Kcomp_P}
satisfies the fixed-point condition, Eq. \ref{eq_dla_dPi_sym}. First,
from Eq. \ref{eq_trPP_ortho},
\begin{equation}
\sum_{l=1}^{N}tr\left[P_{l}P_{i}^{T}\right]I_{N}P_{l}=\lambda KP_{i}.
\end{equation}
$(r+1)$-th diagonal block component of $\sum_{l}\left[P_{l}P_{i}^{T}+P_{i}P_{l}^{T}\right]P_{l}$
is written as
\begin{equation}
\begin{aligned}\left[\sum_{l=1}^{N}\left(P_{l}P_{i}^{T}+P_{i}P_{l}^{T}\right)P_{l}\right]_{(r+1)\text{-th block}} & =\sum_{l=1}^{N}\left(\sum_{k=1}^{K}\sum_{m=1}^{K}W_{l,rK+k}W_{i,rK+m}\left[A_{k}A_{m}^{T}+A_{m}A_{k}^{T}\right]\right)\left(\sum_{n=1}^{N}W_{l,rK+n}A_{n}\right)\\
 & =2\lambda\sum_{k=1}^{K}\sum_{n=1}^{K}W_{i,rK+k}\left(\sum_{l=1}^{N}W_{l,rK+k}W_{l,rK+n}\right)A_{n}\\
 & =2\lambda\left[P_{i}\right]_{(r+1)\text{-th block}}.
\end{aligned}
\label{eq_PlPiT_PiPlT_Pl}
\end{equation}
In addition, we have 
\begin{equation}
\begin{aligned}\sum_{l=1}^{N}\left[P_{l}P_{l}^{T}\right]_{(r+1)\text{-th block}} & =\sum_{i=1}^{N}\left(\sum_{k=1}^{K}W_{l,rK+k}A_{k}\right)\left(\sum_{n=1}^{K}W_{l,rK+n}A_{n}^{T}\right)\\
 & =\sum_{k=1}^{K}\sum_{n=1}^{K}\left[W^{T}W\right]_{rK+k,rK+n}A_{k}A_{n}^{T}=K\lambda I_{K},
\end{aligned}
\label{eq_PlPlt_ortho}
\end{equation}
By combining the equations above, we get
\begin{equation}
\sum_{l=1}^{N}\left(tr\left[P_{l}P_{i}^{T}\right]I_{N}+L\cdot P_{i}P_{l}^{T}+P_{l}P_{i}^{T}\right)P_{l}=\left(\lambda K+2\lambda+\left[L-1\right]\lambda K\right)P_{i}=P_{i}.
\end{equation}
Thus, Eq. \ref{eq_def_ortho_Kcomp_P} indeed satisfies the fixed-point
condition, Eq. \ref{eq_dla_dPi_sym}.

\paragraph{Decoding error of $\boldsymbol{b}$}

Let us next consider the decoding error of $\boldsymbol{b}$, $\ell_{b}$.
Under $P=R$, the error $\ell_{b}$ is written as 
\begin{equation}
\ell_{b}=1-\frac{2}{N}\sum_{i}tr\left[P_{i}P_{i}^{T}\right]+\frac{1}{N}\sum_{i}\sum_{l}\left(tr\left[P_{i}P_{i}^{T}P_{l}P_{l}^{T}\right]+tr\left[P_{l}P_{i}^{T}\left(L\cdot P_{i}P_{l}^{T}+P_{l}P_{i}^{T}\right)\right]\right).
\end{equation}
From Eq. \ref{eq_PlPlt_ortho}, we get
\begin{equation}
\sum_{i=1}^{N}\sum_{l=1}^{N}tr\left[P_{i}P_{i}^{T}P_{l}P_{l}^{T}\right]=tr\left[\left(K\lambda\right)^{2}I_{N}\right]=NK^{2}\lambda^{2}.
\end{equation}
Because the rest of terms are the same with $\ell_{a}$, the error
$\ell_{b}$ also follows 
\begin{equation}
\ell_{b}=1-2K\lambda_{K}+K^{2}\lambda_{K}^{2}+2K\lambda_{K}^{2}+\left(L-1\right)\lambda^{2}K^{2}=\frac{\left(L-1\right)K+2}{LK+2}.
\end{equation}
Moreover, $\left\{ P_{i}\right\} _{i=1}^{N}$ constructed by Eq. \ref{eq_def_ortho_Kcomp_P}
satisfies the fixed-point condition Eq. \ref{eq_dlb_dPi_sym}. From
Eqs. \ref{eq_PlPiT_PiPlT_Pl} and \ref{eq_PlPlt_ortho}, it follows
that
\begin{equation}
\begin{aligned}\sum_{l=1}^{N}\left(L\cdot P_{i}P_{l}^{T}P_{l}+P_{l}P_{i}^{T}P_{l}+P_{l}P_{l}^{T}P_{i}\right) & =\left(L-1\right)P_{i}\sum_{l}P_{l}^{T}P_{l}+\sum_{l}\left(P_{i}P_{l}^{T}+P_{l}P_{i}^{T}\right)P_{l}+\sum_{l}P_{l}P_{l}^{T}P_{i}\\
 & =\left(L-1\right)\lambda KP_{i}+2\lambda P_{i}+\lambda LP_{i}=P_{i}.
\end{aligned}
\end{equation}

\subsection{Extended octonion binding\label{subsec:Extended-Octonion-binding} }

The sparse octonion binding can be naturally extended to $N_{c}>N$
when $N_{c}$ satisfies $N_{c}=dN$ for a positive integer $d$. As
before, we set $N$ to be $N=qK$ for a positive integer $q$. 

Let us focus on the case when $N_{c}<N^{2}/K$ for simplicity. Using
a solution for Eq. \ref{eq_Hurwitz_mat_norm}, $\left\{ A_{i}\right\} _{i=1}^{K}$,
we introduce a family of $N\times N$ matrix $\left\{ B^{\mu\nu}\right\} $
as
\begin{equation}
B^{\mu\nu}\equiv\underbrace{O_{K}\oplus...\oplus O_{K}}_{\nu}\oplus A_{\mu}\oplus O_{K}\oplus...\oplus O_{K}
\end{equation}
for $\mu=1,...,K$ and $\nu=0,...,q-1$. We then construct a family
of $N\times N_{c}$ matrices $\left\{ P_{i}\right\} _{i=1}^{N}$ from
$B$ as
\begin{equation}
P_{i}=\left[B^{\left\lceil i/q\right\rceil ,i\%q},B^{\left\lceil i/q\right\rceil ,(i+1)\%q},...,B^{\left\lceil i/q\right\rceil ,(i+d-1)\%q}\right].
\end{equation}
For instance, if $q=3$ and $d=2$, then 
\begin{equation}
\begin{aligned}P_{1}=\left(\begin{array}{cccccc}
A_{1} & O & O & O & O & O\\
O & O & O & O & A_{1} & O\\
O & O & O & O & O & O
\end{array}\right),\quad & P_{2}=\left(\begin{array}{cccccc}
O & O & O & O & O & O\\
O & A_{1} & O & O & O & O\\
O & O & O & O & O & A_{1}
\end{array}\right),\\
P_{3}=\left(\begin{array}{cccccc}
O & O & O & A_{1} & O & O\\
O & O & O & O & O & O\\
O & O & A_{1} & O & O & O
\end{array}\right),\quad & P_{4}=\left(\begin{array}{cccccc}
A_{2} & O & O & O & O & O\\
O & O & O & O & A_{2} & O\\
O & O & O & O & O & O
\end{array}\right),...
\end{aligned}
\end{equation}
Let us estimate the error under this binding method. From the definition,
$P_{i}P_{l}^{T}$ is written as
\begin{equation}
P_{i}P_{l}^{T}=\sum_{r=0}^{d-1}\left[(i+r)\%q=(l+r)\%q\right]_{+}\underbrace{O_{K}\oplus...\oplus O_{K}}_{(i+r)\%q}\oplus A_{\left\lceil i/q\right\rceil }A_{\left\lceil l/q\right\rceil }^{T}\oplus O_{K}\oplus...\oplus O_{K},
\end{equation}
where $\left[x\right]_{+}$ is an indicator function that returns
1 if $x$ is true, and returns 0 if false. Thus, $tr\left[P_{i}P_{i}^{T}\right]=d\lambda_{K}K$.
This means that, in order to satisfy Eq. \ref{eq_PQ_tr_constraint},
the scaling factor $\lambda_{K}$ of $A_{k}$ in Eq. \ref{eq_Hurwitz_mat_norm}
needs to be $\lambda_{K}=1/(dK)$. The dominant term of the error
becomes
\begin{equation}
\begin{aligned}\frac{L}{N}\sum_{i=1}^{N}\sum_{l=1}^{N}tr\left[P_{l}P_{i}^{T}P_{i}P_{l}^{T}\right] & =\frac{L}{N}\sum_{i=1}^{N}\sum_{l=1}^{N}\sum_{r=0}^{d-1}\left[(i+r)\%q=(l+r)\%q\right]_{+}tr\left[A_{\left\lceil l/q\right\rceil }A_{\left\lceil i/q\right\rceil }^{T}A_{\left\lceil i/q\right\rceil }A_{\left\lceil l/q\right\rceil }^{T}\right]\\
 & =\frac{L}{N}\sum_{r=0}^{d-1}\sum_{m=1}^{q}\sum_{k=1}^{K}\sum_{m^{\prime}=1}^{q}\sum_{k^{\prime}=1}^{K}\left[\left([k-1]q+m\right)\%q=\left([k^{\prime}-1]q+m^{\prime}\right)\%q\right]_{+}tr\left[A_{k^{\prime}}A_{k}^{T}A_{k}A_{k^{\prime}}^{T}\right]\\
 & =\frac{L}{N}\sum_{r=0}^{d-1}\sum_{m=1}^{q}\sum_{m^{\prime}=1}^{q}\left[m\%q=m^{\prime}\%q\right]_{+}\left(\lambda_{K}K\right)^{2}K\\
 & =\frac{L}{N}\frac{dN}{K}\lambda_{K}^{2}K^{3}=\frac{LN}{N_{c}}
\end{aligned}
\end{equation}
Therefore, the dominant term is the same with the lower bound obtained
in Appendix \ref{subsec:Lower-bound}. Calculating the rest of terms
in a similar manner, we get 
\begin{equation}
\ell_{a}=\frac{N}{N_{c}}\left(L-1+\frac{2}{K}\right).\label{eq_la_extended_octonions}
\end{equation}

\subsection{Construction of higher-order sparse $K$-compositional bindings}

In the simulations depicted in Figs 4,5,7-9, we constructed sparse
$K$-compositional bindings by using a python library for the Cayley-Dickson
construction, developed by Dr. Travis Hoppe (https://github.com/thoppe/Cayley-Dickson).
Source codes for the simulations are available at https://github.com/nhiratani/quadratic\_binding. 

\section{Tensor-HRR bindings\label{sec:Tensor-Holographic-bindings}}

Below, we review two commonly used binding mechanisms: holographic
reduced representation (HRR) \citep{plate1995holographic,plate1997common,nickel2016holographic}
and tensor product representation \citep{smolensky1990tensor,smolensky2014optimization}.
Subsequently, we introduce a binding that morphs from HRR to the tensor
product representation as you change the vector length of the representation. 

\subsection{Holographic reduced representation (HRR)\label{subsec:HRR}}

Under HRR, the length of the composition vector $\boldsymbol{c}$
is the same with that of $\boldsymbol{a}$ and $\boldsymbol{b}$ ($N_{c}=N$),
and the $k$-th element of binding $\psi$ is constructed by
\begin{equation}
\psi_{k}(\boldsymbol{a}_{\mu},\boldsymbol{b}_{\mu})=\sum_{i=1}^{N}a_{i}^{\mu}b_{[k-i]_{N}}^{\mu}\;\text{for}\;k=1,...,N\label{eq_holog_binding}
\end{equation}
where $[k-i]_{N}\equiv k-i\;(\textrm{mod}.\;N)$, and $a_{i}^{\mu}$
is the $i$-th element of vector $\boldsymbol{a}_{\mu}$. Given $\boldsymbol{c}=\sum_{\mu=1}^{L}\psi(\boldsymbol{a}_{\mu},\boldsymbol{b}_{\mu})$,
we can unbind $\boldsymbol{a}_{1}$ from $\boldsymbol{c}$ using a
query $\boldsymbol{b}_{1}$ as
\begin{equation}
\widehat{a}_{i}^{1}=\frac{1}{\left\Vert \boldsymbol{b}_{1}\right\Vert ^{2}}\sum_{j=1}^{N}c_{[j+i]_{N}}b_{j}^{1}.\label{eq_holog_unbinding}
\end{equation}
Because $\boldsymbol{c}$ is rewritten as 
\begin{equation}
c_{k}=\sum_{\mu=1}^{L}\sum_{i=1}^{N}a_{i}^{\mu}b_{[k-i]_{N}}^{\mu}=\sum_{\mu=1}^{L}\sum_{i=1}^{N}\sum_{j=1}^{N}a_{i}^{\mu}b_{j}^{\mu}\delta_{[i+j]_{N},k},
\end{equation}
this is a quadratic binding with
\begin{equation}
P_{ijk}=\delta_{[i+j]_{N},k}.
\end{equation}
Similarly, if the amplitude of $\boldsymbol{a}_{\mu}$ and $\boldsymbol{b}_{\mu}$
are normalized as $\left\Vert \boldsymbol{a}_{\mu}\right\Vert ^{2}=\left\Vert \boldsymbol{b}_{\mu}\right\Vert ^{2}=N$,
unbinding of $\boldsymbol{a}$ and $\boldsymbol{b}$ are given as
\begin{equation}
Q_{ijk}=R_{ijk}=\frac{1}{N}\delta_{[i+j]_{N},k}.
\end{equation}
Alternatively, by moving the half of the normalization factor to the
binding operator, we can rewrite $P,Q,R$ as
\begin{equation}
P_{ijk}=Q_{ijk}=R_{ijk}=\frac{1}{\sqrt{N}}\delta_{[i+j]_{N},k}.\label{eq_hrr_quadratic}
\end{equation}
Under both normalizations, the amplitude of the recovered signal becomes
the same with the original signal amplitude (see Eq. \ref{eq_unbinding_amplitude}).
However, this normalization does not necessarily minimize the mean-squared
error $\ell_{a}$ and $\ell_{b}$. To see this, let us define $P_{ijk}=Q_{ijk}=R_{ijk}=\sqrt{\lambda}\delta_{[i+j]_{N},k}$
where $\lambda$ is a scaling factor. Then, from Eq. \ref{eq_la_L1_2},
assuming that $N$ is an even number, the loss $\ell_{a}$ becomes
\begin{equation}
\ell_{a}=1-2N\lambda+\left(\left(L+1\right)N+2\right)N\lambda^{2}.
\end{equation}
Thus, the loss is minimized at $\lambda=\frac{1}{(L+1)N+2}$, under
which the loss follows $\ell_{a}=\frac{LN+2}{\left(L+1\right)N+2}$.
In particular, when only one pair is bound to the composition ($L=1$),
we get $\ell_{a}=\frac{N+2}{2(N+1)}$ (blue line in Fig. 4A), and
taking the large $N$ limit, we obtain $\ell_{a}=\frac{1}{2}$ (gray
dashed line in Fig. 2A and B). If we instead set $\lambda=\frac{1}{N}$
to make the decoding unbiased, the loss follows $\ell_{a}=1+\frac{2}{N}$
under $L=1$ (blue line in Fig. 4C). The loss $\ell_{b}$ becomes
the same due to the symmetry between $\boldsymbol{a}$ and $\boldsymbol{b}$
under HRR (Eq. \ref{eq_hrr_quadratic} is invariant against $i\leftrightarrow j$).

It should be noted that HRR does not satisfy the fixed-point conditions
(Eqs. \ref{eq_dla_dPi_sym} and \ref{eq_dlb_dPi_sym}) under a finite
$N$ regardless of the choice of the scaling factor. This is because
the right-hand side of Eq. \ref{eq_dla_dp_sym} becomes
\begin{equation}
\sum_{l=1}^{N}\left(tr\left[P_{l}P_{i}^{T}\right]I_{N}+L\cdot P_{i}P_{l}^{T}+P_{l}P_{i}^{T}\right)P_{l}=\lambda\left(L+1\right)NP_{i}+\lambda P_{i}^{res},
\end{equation}
where $\left[P_{i}^{res}\right]_{jk}\equiv\sqrt{\lambda}\sum_{l}\delta_{[2l]_{N},[i-j+k]_{N}}$.
Nonetheless, the fact that $P=Q$ is satisfied under HRR is consistent
with the condition on the optimal $Q$ at $L\gg1$ limit (Eq. \ref{eq_optQi_Lagrange}).
Under HRR, 
\begin{equation}
\left[\sum_{l=1}^{N}P_{l}P_{l}^{T}\right]_{jm}=\sum_{l=1}^{N}\sum_{k=1}^{N}P_{ljk}P_{lmk}=\frac{1}{N}\sum_{l=1}^{N}\delta_{[l+j]_{N},[l+m]_{N}}=\delta_{jm}.
\end{equation}
Thus, for a given $P$, to minimize the Lagrangian, $Q$ needs to
satisfy $Q_{i}=\lambda_{i}P_{i}$ for $i=1,...,N$. This result supports
the optimality of unbinding by circular correlation given a binding
by circular convolution at $L\gg1$. From the symmetry between $\boldsymbol{a}$
and $\boldsymbol{b}$ ($P_{ijk}=P_{jik}$), we expect $R_{i}=\lambda_{i}P_{i}$
to be the optimal too at $L\gg1$. 

\subsection{Tensor product representation\label{subsec:TPR}}

In the tensor product representation, $S=\left\{ \left(\boldsymbol{a}_{\mu},\boldsymbol{b}_{\mu}\right)\right\} _{\mu=1}^{L}$
is represented by a $N\times N$ matrix $\boldsymbol{C}$:
\begin{equation}
\boldsymbol{C}=\sum_{\mu=1}^{L}\boldsymbol{a}_{\mu}\boldsymbol{b}_{\mu}^{T}
\end{equation}
Alternatively, we can consider $\boldsymbol{C}$ as a length $N_{c}=N^{2}$
vector $\boldsymbol{c}=\text{Vec}[\boldsymbol{C}]$. Given $\boldsymbol{C}$
and a query $\boldsymbol{b}_{1}$, unbinding of $\boldsymbol{a}_{1}$
is done by
\begin{equation}
\widehat{\boldsymbol{a}}_{1}=\frac{1}{\left\Vert \boldsymbol{b}_{1}\right\Vert ^{2}}\boldsymbol{C}\boldsymbol{b}_{1}
\end{equation}
This unbinding is lossless if $L=1$ because $\frac{1}{\left\Vert \boldsymbol{b}_{1}\right\Vert ^{2}}\boldsymbol{a}_{1}\boldsymbol{b}_{1}^{T}\boldsymbol{b}_{1}=\boldsymbol{a}_{1}$.
The tensor product representation is also an example of quadratic
binding family in which, assuming $\left\Vert \boldsymbol{a}_{\mu}\right\Vert ^{2}=\left\Vert \boldsymbol{b}_{\mu}\right\Vert ^{2}=N$,
the tensors $P,Q,R$ are set to
\begin{equation}
P_{ijk}=Q_{ijk}=R_{ijk}=\frac{1}{\sqrt{N}}\delta_{k,(iN+j)}.\label{eq_tensor_quadratic}
\end{equation}
Notably, $P=Q=R$ is satisfied in the tensor product representation
too. 

\subsection{Tensor-HRR morphing\label{subsec:Tensor-HRR-morphing}}

For $N_{c}=dN$ with $d=1,2,...,N$, we define tensor-HRR binding
as
\begin{equation}
P_{ijk}=Q_{ijk}=R_{ijk}=\frac{1}{\sqrt{N}}\delta_{[id+j]_{dN},k}
\end{equation}
At $N_{c}=N$ ($d=1$), this is the same with HRR (Eq. \ref{eq_hrr_quadratic}),
whereas at $N_{c}=N^{2}$, $\delta_{[iN+j]_{N^{2}},k}=\delta_{(iN+j),k}$,
thus it becomes the tensor-product binding (Eq. \ref{eq_tensor_quadratic}).
Noticing that $M$ (Eq. \ref{eq_def_tensorM}) is written as 
\begin{equation}
M_{mj}^{li}=\frac{1}{N}\delta_{[id+j]_{dN},[ld+m]_{dN}},\label{eq_tensorM_tenHRR}
\end{equation}
 unbinding of $\boldsymbol{a}_{1}$ indeed yields
\begin{equation}
\begin{array}{rl}
\left\langle \widehat{a}_{i}^{1}\right\rangle _{p(b)} & =\left\langle \sum_{j}\sum_{k}P_{ijk}b_{j}^{1}\sum_{\mu}\sum_{l}\sum_{m}P_{lmk}a_{l}^{\mu}b_{m}^{\mu}\right\rangle _{p(b)}\\
 & =\left\langle \sum_{j}\sum_{l}M_{jj}^{li}\left(b_{j}^{1}\right)^{2}a_{l}^{1}\right\rangle _{p(b)}\\
 & =\frac{1}{N}\sum_{j}\sum_{l}\delta_{[id+j]_{dN},[ld+j]_{dN}}a_{l}^{1}\\
 & =a_{i}^{1}.
\end{array}
\end{equation}
In the last line, we used 
\begin{equation}
id+j\equiv ld+j\;(mod.\;dN)\Leftrightarrow(i-l)d\equiv0\;(mod.\;dN)\Leftrightarrow i=l,
\end{equation}
for $i,l=1,...,N$. The decoding error (Eq. \ref{eq_la_full}) under
this binding is estimated as below. First, from Eq. \ref{eq_tensorM_tenHRR}
\begin{equation}
1-\frac{2}{N}\sum_{i}\sum_{j}M_{jj}^{ii}+\frac{1}{N}\sum_{i}\sum_{l}\left(\sum_{j}M_{jj}^{li}\right)^{2}=1-2+1=0.
\end{equation}
On the other hand, under $N\gg1$, the noise term is given as
\begin{equation}
\begin{aligned}\frac{1}{N}\sum_{i}\sum_{l}\sum_{j}\sum_{m}M_{mj}^{li}\left(LM_{mj}^{li}+M_{mj}^{il}\right) & =\frac{1}{N^{3}}\sum_{i}\sum_{l}\sum_{j}\sum_{m}\left(L\delta_{[id+j]_{dN},[ld+m]_{dN}}+\delta_{[id+j]_{dN},[ld+m]_{dN}}\delta_{[ld+j]_{dN},[id+m]_{dN}}\right)\\
 & \approx\frac{L}{d}+\frac{1}{N}.
\end{aligned}
\end{equation}
The last line follows under a large $N$, because for randomly sampled
integers $1\leq i,l,j,m\leq N$, 
\begin{equation}
\Pr\left[\delta_{[id+j]_{dN},[ld+m]_{dN}}=1\right]=\frac{1}{dN}.
\end{equation}
Combining the terms above, we get 
\begin{equation}
\ell_{a}\approx\frac{LN}{N_{c}}+\frac{1}{N}.\label{eq_la_tenHRR}
\end{equation}

\subsection*{Data availability }

Source code is available at https://github.com/nhiratani/quadratic\_binding.

\subsection*{Acknowledgements }

This work has been supported by the Swartz Foundation. 

\bibliographystyle{apalike}
\bibliography{refs}

\end{document}